\documentclass[submission, Phys]{SciPost}

\usepackage[utf8]{inputenc} % input encodings (allow UTF-8 input)
\usepackage[T1]{fontenc} 	% selecting font encodings (use 8-bit T1 fonts)
\usepackage[english]{babel} % multilingual support (English language/hyphenation)

\usepackage[bitstream-charter]{mathdesign}
\usepackage{geometry} 		% flexible and complete interface to document dimensions
\usepackage{amsmath} 		% math package (American Mathematical Society)
\usepackage{mathtools} 		% math package (fixes various deficiencies of amsmath)
\usepackage{float} 			% floating objects such as figures and tables
\usepackage{graphicx} 		% enhanced support for graphics
\usepackage{tabularx} 		% tabulars with adjustable-width columns
\usepackage{booktabs} 		% professional-quality tables
\usepackage{color, xcolor} 	% foreground and background colour management
\usepackage{pdfpages} 		% inclusion of external multi-page PDF documents
\usepackage{extarrows} 		% extra arrows beyond those provided in amsmath
\usepackage{multirow} 		% create tabular cells spanning multiple rows
\usepackage{multicol} 		% intermix single and multiple columns
\usepackage{enumitem} 		% control layout of itemize, enumerate, description
\usepackage{xspace} 		% define commands that appear not to eat spaces
\usepackage{stackrel} 		% enhancement to the \stackrel command
\usepackage{tikz} 			% create PostScript and PDF graphics
\usepackage{braket} 		% Dirac bra-ket notation
\usepackage{bm} 			% access bold symbols in maths mode
\usepackage{tensor} 		% typeset tensors (tensor-style super- and subscripts)
\usepackage{slashed} 		% slash through characters (Feynman slash notation)
\usepackage{siunitx} 		% SI units package (typesetting values with units)
\usepackage{lastpage} 		% reference last page
\usepackage{cite} 			% improved citation handling
\usepackage[normalem]{ulem} % package for underlining
\usepackage{fontawesome} 	% access to web-related icons
\usepackage{tocloft} 		% control over the typography of the TOC, LOF, and LOT
\usepackage{titlesec} 		% interface to sectioning commands (various title styles)
\usepackage{doi} 			% create correct hyperlinks for DOI numbers
\usepackage{hyperref} 		% hypertext links (handel cross-referencing commands)
\usepackage[most]{tcolorbox} 					% coloured and framed text boxes
\usepackage[nameinlink, capitalize]{cleveref} 	% intelligent cross-referencing
\usepackage[nottoc, notlot, notlof]{tocbibind} 	% add references/index/contents to TOC
\usepackage[ruled, vlined]{algorithm2e} 		% floating algorithm environment
\usepackage{makecell}
\usepackage[makeroom]{cancel}
\usepackage{feynmf}

\makeatletter
\def\BState{\State\hskip-\ALG@thistlm}
\makeatother

\makeatletter
\@ifundefined{pdfoutput}{}{\DeclareGraphicsRule{*}{mps}{*}{}}
\makeatother

\makeatletter
\DeclareRobustCommand*{\bfseries}{%
   \not@math@alphabet\bfseries\mathbf
   \fontseries\bfdefault\selectfont
   \boldmath
}
\makeatother

\hypersetup{
	pdftitle={Generative Amplification with Surrogate Monte Carlo},
	pdfauthor={Bahl et al.},
	colorlinks=true, 			% false: boxed links, true: colored links
	linkcolor={red!50!black}, 	% color of internal links (sections, pages, etc.)
	citecolor={blue!50!black}, 	% color of citation links (links to bibliography)
	urlcolor={blue!80!black} 	% color of URL links (external links)
} 
\DeclareSymbolFont{usualmathcal}{OMS}{cmsy}{m}{n}
\DeclareSymbolFontAlphabet{\mathcal}{usualmathcal}

\SetArgSty{textnormal}
\SetKwComment{Comment}{{\small\#}~}{}
\SetCommentSty{mycommfont}

\setitemize{itemsep=0pt, parsep=0pt} 				% adjust itemize environment
\setenumerate{itemsep=0pt, parsep=0pt} 				% adjust enumerate environment
\setitemize{itemsep=2pt,topsep=2pt,parsep=0pt,partopsep=0pt,leftmargin=*}
\setenumerate{itemsep=0pt,topsep=2pt,parsep=0pt,partopsep=0pt,labelindent=3pt,leftmargin=*}
\usepackage{amsmath}
 
\usepackage{amsthm} 		% math package (typesetting theorems using AMS style)
\theoremstyle{definition}
 
\definecolor{red_cb}{HTML}{e41a1c}
\definecolor{blue_cb}{HTML}{377eb8}
\definecolor{green_cb}{HTML}{4daf4a}
\definecolor{purple_cb}{HTML}{984ea3}
\definecolor{orange_cb}{HTML}{ff7f00}

\definecolor{EmeraldGreen}{HTML}{1ea78d}
\definecolor{EnglishRed}{HTML}{b02427}
\hypersetup{colorlinks=true,urlcolor=EmeraldGreen,citecolor=EmeraldGreen,linkcolor=EnglishRed}

\newcommand{\ie}{\text{i.e.}\;}

\newcommand{\eqcomma}{\quad\text{,}} 	% equation comma
\newcommand{\eqperiod}{\quad\text{.}} 	% equation period

\newcommand{\Nholdout}{n_\text{holdout}}
\newcommand{\Neff}{n_\text{eff}}
\newcommand{\Nequiv}{n_\text{equiv}}
\newcommand{\Ntrain}{n_\text{train}}

\newcommand{\Ngen}{n_\text{gen}}

\newcommand{\pgen}{p_\text{surr}}
\newcommand{\ptrue}{p_\text{true}}

\newcommand{\XLangle}{\Bigl\langle}
\newcommand{\XRangle}{\Bigr\rangle}

\newcommand{\qqquad}{\qquad\quad}
\newcommand{\qqqquad}{\qquad\qquad}

\newcommand\one{\leavevmode\hbox{\small1\normalsize\kern-.33em1}}
\newcommand{\really}{\stackrel{!}{=}}

\newcommand{\loss}{\mathcal{L}} 	% loss value

\newcommand{\sherpa}{\textsc{Sherpa}\xspace}

\newcommand{\madnis}{\textsc{MadNIS}\xspace}

\newcommand{\arXiv}[2][]{%
	\ifthenelse{\equal{#1}{}}%
	{\href{http://arxiv.org/abs/#2}{arXiv:#2}}%
	{\href{http://arxiv.org/abs/#2}{arXiv:#2~[#1]}}}

\newcommand{\gev}{\text{GeV}}

\def\slashchar#1{\setbox0=\hbox{$#1$}           % set a box for #1
   \dimen0=\wd0                                 % and get its size
   \setbox1=\hbox{/} \dimen1=\wd1               % get size of /
   \ifdim\dimen0>\dimen1                        % #1 is bigger
      \rlap{\hbox to \dimen0{\hfil/\hfil}}      % so center / in box
      #1                                        % and print #1
   \else                                        % / is bigger
      \rlap{\hbox to \dimen1{\hfil$#1$\hfil}}   % so center #1
      /                                         % and print /
   \fi}

\newcommand{\tikznode}[2]{%
\ifmmode%
\tikz[remember picture,baseline=(#1.base),inner sep=0pt] \node (#1) {$#2$};%
\else
\tikz[remember picture,baseline=(#1.base),inner sep=0pt] \node (#1) {#2};%
\fi}

\def\mathswitchr#1{\relax\ifmmode{\mathrm{#1}}\else$\mathrm{#1}$\xspace\fi}
\def\mathswitch#1{\relax\ifmmode#1\else$#1$\xspace\fi}

\graphicspath{{./figs/}}

\begin{document}

\begin{center}{\Large \textbf{
%Extrapolating Monte Carlos with Amplitude Surrogates}}
Generative Amplification with Surrogate Monte Carlo}}
\end{center}

\begin{center}
  Henning Bahl\textsuperscript{1},
  Tilman Plehn\textsuperscript{1,2},
  and Rebecca Revelli\textsuperscript{1}
\end{center}

\begin{center}
  {\bf 1} Institut f\"ur Theoretische Physik, Universit\"at Heidelberg, Germany \\
  {\bf 2} Interdisciplinary Center for Scientific Computing (IWR), Universit\"at Heidelberg, Germany
\end{center}

\begin{center}
\today
\end{center}

% For convenience during refereeing: line numbers
%\linenumbers

\section*{Abstract}
{\bf
  Amplitude surrogates for LHC simulations build on generative amplification, the fact that a surrogate trained on an expensive and small training dataset describes the smooth amplitude more precisely than the training data does. Applying techniques developed for generative networks, we quantify this amplification for gluon-associated $Z$ production. Significant amplification appears in sparsely populated kinematic tails, where it matters most. Our results show how generative amplification from surrogate Monte Carlo far outperforms the density estimation in current generative networks. 
}

% TODO: include a table of contents (optional)
\vspace{10pt}
\noindent\rule{\textwidth}{1pt}
\tableofcontents\thispagestyle{fancy}
\noindent\rule{\textwidth}{1pt}
\vspace{10pt}

\clearpage
%%%%%%%%%%%%%%%%%%%%%%%%%%%%%%%%%%%%%%%%%%%%%%%%%%%
\section{Introduction}
\label{sec:intro}

The great success of LHC physics is the development of the first precision-hadron collider program. It critically relies on comparing precisely measured rates with equally precise first-principles predictions. The rapidly growing dataset drives these precision requirements through ever-decreasing statistical uncertainties. After the high-luminosity upgrade, many LHC analyses will be limited not by data but by our ability to predict the corresponding events with sufficient precision and in sufficient numbers.

LHC simulations follow a factorized chain of first-principles predictions: a hard scattering process described by perturbative QCD, the sampling of the corresponding phase space, jet radiation and parton shower, and finally hadronization and hadron decays. For the first step, the statistical power of the coming LHC runs forces us to include higher multiplicities and higher perturbative orders. This additional cost is not covered by the usual prediction of the LHC computing budget and needs to be absorbed through improved numerical methods.

Modern machine learning offers two complementary solution strategies~\cite{Butter:2022rso,Plehn:2022ftl}. The first replaces the entire simulation chain by an end-to-end generative model. Generative adversarial networks~\cite{Otten:2019hhl,Hashemi:2019fkn,DiSipio:2019imz,Butter:2019cae,Alanazi:2020klf,Choi:2021sku}, normalizing flows~\cite{Gao:2020zvv,Stienen:2020gns,Verheyen:2022tov} with controlled uncertainties~\cite{Bellagente:2021yyh,Butter:2021csz}, and diffusion networks and transformer architectures~\cite{Butter:2023fov,Heimel:2023mvw} can extract the phase-space density from a training sample of events. The networks then generate new events orders of magnitude faster than the classical chain. Because the density estimation in the generative network training assumes a smooth density, the generated sample can describe the truth better than its finite training data, the so-called amplification effect~\cite{Butter:2020qhk,Bieringer:2022cbs,Bieringer:2024nbc,Bahl:2025ryd}.

The second strategy keeps the factorized chain intact and only accelerates its expensive modules. The matrix element can be replaced by a fast network surrogate trained on a limited set of exact evaluations that then predicts the amplitude at negligible cost~\cite{Bishara:2019iwh,Badger:2020uow,Aylett-Bullock:2021hmo,Maitre:2021uaa,Danziger:2021eeg,Winterhalder:2021ngy,Badger:2022hwf,Janssen:2023ahv,Maitre:2023dqz,Brehmer:2024yqw,Breso-Pla:2024pda,Bahl:2024gyt,Bahl:2025xvx,Herrmann:2025nnz,Favaro:2025pgz,Villadamigo:2025our,Beccatini:2025tpk,Bahl:2026qaf,Bahl:2026jvt,Dubey:2026cts}. Neural important sampling~\cite{Bendavid:2017zhk,Klimek:2018mza,Chen:2020nfb,Gao:2020vdv,Deutschmann:2024lml} has been successfully applied at leading order (LO) using \madnis~\cite{Heimel:2022wyj,Heimel:2023ngj,Heimel:2024wph,Heimel:2026hgp,Heimel:2026cxh} and \sherpa~\cite{Gao:2020zvv, Bothmann:2020ywa, Bothmann:2025lwg}, going to NLO~\cite{Gao:2020zvv,DeCrescenzo:2026tsp} and NNLO accuracy in multi-jet final states~\cite{Janssen:2025zke}. The network precision can be improved through equivariant transformers encoding the Lorentz symmetry~\cite{Brehmer:2024yqw,Favaro:2025pgz,Petitjean:2025zjf} and interpolation or factorization-aware constructions for high multiplicities and full color~\cite{Breso-Pla:2024pda,Herrmann:2025nnz,Villadamigo:2025our,Beccatini:2025tpk}. Modern amplitude surrogates are accurate below the per-mille level and come with calibrated learned uncertainties~\cite{Bahl:2024gyt,Bahl:2025xvx, Beccatini:2025tpk,Bahl:2026qaf,Bahl:2026jvt,Dubey:2026cts}.

The amplification question carries over from generative networks to surrogate Monte Carlo. Again, we can ask if an amplitude surrogate, trained on a limited expensive dataset, delivers the statistical power of a much larger true sample. The challenge for LHC predictions is sparsely populated tails, where the truth simulation runs out of events while the amplitude provides a smooth interpolation or even extrapolation. We adapt our amplification framework~\cite{Bahl:2025ryd} to amplitude surrogates and apply it to an uncertainty-aware L-GATr surrogate~\cite{Brehmer:2024yqw,Petitjean:2025zjf,Favaro:2026amj} for $Z+$gluons production in the kinematic tails.

In Section~\ref{sec:concepts} we introduce a reweighting setup to describe surrogate amplification and define the averaging and differential amplification measures.
In Section~\ref{sec:results} we present our results for the $Z+$gluons benchmark amplitudes. For $Z + g$ and for $Z+4g$ production we see that in particular in the kinematic tails the surrogate amplitude Monte Carlo generates massive statistical amplification. We conclude in Section~\ref{sec:outlook}. In a series of Appendices we provide additional information, including results for the $Z + 2g$ and $Z+ 3g$ processes.

%%%%%%%%%%%%%%%%%%%%%%%%%%%%%%%%%%%%%%%%%%%%%%%%%%%
\section{Amplification for amplitude surrogates}
\label{sec:concepts}

When we train a generative network, it has to implicitly or explicitly encode the phase space density underlying the training data,
\begin{align}
   \pgen (x) \approx \ptrue (x) \; .
\label{eq:train1}
\end{align}
We expect the generated dataset to describe the true distribution better than the statistically limited training data does. Such an amplification can be driven by the limited resolution of the family of smooth functions that the network uses to encode the density. It can be viewed as an extrapolation in resolution. For LHC transition amplitudes, we know that there are no finer structures than intermediate mass peaks with GeV-scale widths. 

We use a first-principle training dataset with $\Ntrain$ data points, $D_\text{true}^{\Ntrain}$. The network training in Eq.\eqref{eq:train1} then follows
\begin{align}
  \pgen(x) \approx D_\text{true}^{\Ntrain}  \sim \ptrue(x) \; .
\label{eq:train2}
\end{align}
Changing notation compared to Ref.~\cite{Bahl:2025ryd}, we denote the surrogate-generated dataset of size $\Ngen$ as $D_\text{surr}^{\Ngen}$. The quality of the generative surrogate is determined by the relation
\begin{align}
  D_\text{surr}^{\Ngen} \Bigg|_{\Ngen \to \infty}
  \sim \pgen(x) 
  \stackrel{?}{=} \ptrue(x) \; .
    \label{eq:amp_define}
\end{align}
To analyze it, we need a metric for the agreement of a dataset with a probability distribution, $M[D, p(x)]$. We then define the effective number of samples $\Nequiv$ implicitly, such that an assumed-to-be-true dataset approximates $\ptrue(x)$ as well as an infinitely large dataset following the approximately learned $\pgen(x)$,
\begin{align}
    M\left[D_\text{true}^{\Ngen},\ptrue \right] \Bigg|_{\Ngen = \Nequiv} \really
    M\left[ D_\text{surr}^{\Ngen}, \ptrue \right] \Bigg|_{\Ngen \to \infty}
 \; .
  \label{eq:amplification_core}
\end{align}
If we consider the deviation on the left-hand side as statistical and the deviation on the right-hand side as model systematics, $\Nequiv$ is the maximum number of points for which the statistical deviation exceeds the limiting model systematics. This defines the amplification factor
\begin{align}
    G = \frac{\Nequiv}{\Ntrain} \; . 
\end{align}
This amplification argument holds for implicitly learned phase space densities in generative networks as for explicitly learned Monte Carlo surrogates.

%%%%%%%%%%%%%%%
\subsection{Averaging amplification factor}
\label{sec:reweight}

For the true fiducial cross-section over $V$, or a kinematic bin, we integrate
\begin{align}
  I(\ptrue) = \int_{V}\!\text{d}x\,\ptrue(x) 
            = \int_{V}\!\text{d}x\, F(x) \; |\mathcal{M}|^2_\text{true}(x) \;.
\label{eq:itrue}
\end{align}
The known proportionality factor $F(x)$ will drop out in the relevant ratios below. As we know the surrogate amplitude explicitly, we replace the generation of surrogate events with a convenient, equivalent reweighting of a test dataset of true amplitudes,
\begin{align}
  w_i
  = \frac{|\mathcal{M}|_\text{surr}^2(x_i)}{|\mathcal{M}|^2_\text{true}(x_i)}
  = \frac{\pgen(x_i)}{\ptrue(x_i)} \;.
  \label{eq:weight}
\end{align}
This allows us to estimate $I(\ptrue)$, but from the surrogate dataset $D_\text{surr}^{\Ngen}$,
\begin{align}
     \bar{I}(D_\text{surr}^{\Ngen}) =\frac{w_\text{bin}}{w_\text{tot}}\equiv \frac{\sum_i\,\mathbf{1}_{x_i\in V} w_i}{\sum_i\,w_i} 
     \qquad \text{with} \qquad
     \mathbf{1}_{x \in V} =
     \begin{cases}
       1 & x \in V \\0 & \text{else}
     \end{cases} \; .
\label{eq:local_fraction_definition}
\end{align}
When comparing the estimate $ \bar{I}(D_\text{surr}^{\Ngen})$ to the true $I(\ptrue)$ we encounter a statistical uncertainty and a systematic model uncertainty.

%%%%%%
\subsubsection*{Statistical uncertainty}

If we know the weights, we can estimate the statistical uncertainty by treating $w_\text{bin}$ and $w_\text{tot}$ as random sums and applying first-order error propagation, 
\begin{align}
  \frac{\partial \bar{I}}{\partial w_\text{bin}}
  = \frac{1}{w_\text{tot}}
  \qquad \text{and} \qquad
  \frac{\partial \bar{I}}{\partial w_\text{tot}}
  = - \frac{\bar{I}}{w_\text{tot}} \;.
\end{align}
The necessary variances are
\begin{align}
  \text{Var}(w_\text{bin})=\sum_i \mathbf{1}_{x_i\in V} w_i^2
  \qqquad
  \text{Var}(w_\text{tot})=\sum_i w_i^2
  \qqquad
  \text{Cov}(w_\text{bin},w_\text{tot})=\text{Var}(w_\text{bin})\;,
\end{align}
where the covariance identity follows from $w_\text{tot}=w_\text{bin}+w_\text{not-bin}$ with $w_\text{not-bin}$ independent of $w_\text{bin}$. The statistical variance becomes 
\begin{align}
  \sigma_\text{stat}^2 & = \frac{\text{Var}(w_\text{bin})+\bar{I}^2\,\text{Var}(w_\text{tot})-2\bar{I}\,\text{Cov}(w_\text{bin},w_\text{tot})}{w_\text{tot}^2} \notag \\
  & = \frac{(1-2\bar{I})\sum_i \mathbf{1}_{x_i\in V} w_i^2+\bar{I}^2\sum_i w_i^2}{\bigl(\sum_i w_i\bigr)^2} \;.
  \label{eq:sigmastat}
\end{align}
For weighted events the relevant sample size is given by the Kish effective sample size
\begin{align}
  \Neff = \frac{\bigl(\sum_i w_i\bigr)^2}{\sum_i w_i^2} \; .
  \label{eq:neff}
\end{align}
It agrees with the number of events for uniform weights. We can use it to simplify Eq.\eqref{eq:sigmastat}. If the event weight distribution is uncorrelated with the selection of $V$, we expect
\begin{align}
  \text{Var}(w_\text{bin})= \sum_i\mathbf{1}_{x_i\in V}w_i^2\;\simeq\;\bar{I}\sum_iw_i^2\;.
  \label{eq:weight_bin_approx}
\end{align}
This identity is exact for uniform weights. For the statistical variance, it gives us
\begin{align}
  \sigma_\text{stat}^2
  \;\simeq\;\bar{I}(1-\bar{I})\,\frac{\sum_iw_i^2}{\bigl(\sum_iw_i\bigr)^2}
  \;=\;\frac{\bar{I}(1-\bar{I})}{\Neff}\;.
  \label{eq:sigmastat_neff}
\end{align}
This approximation breaks down for low-density bins populated by a few large-weight events. This motivates the targeted event sampling described below.

%%%%%%
\subsubsection*{Systematic model uncertainty}

If we know the surrogate uncertainty, the uncertainty on the weight is 
\begin{align}
  \sigma_{w,i} 
  \equiv \frac{\sigma_{\pgen}(x_i)}{\ptrue(x_i)}
  = \frac{\sigma_{|\mathcal{M}|^2_\text{surr}}(x_i)}{|\mathcal{M}_\text{true}|^2(x_i)} \;.
\label{eq:def_weights}
\end{align}
We can use it to estimate the systematic model uncertainty on $\bar{I}$ using error propagation,
\begin{align}
  \sigma_\text{model}^2 & = \sum_i \left(\frac{\partial \bar{I}}{\partial w_i}\right)^2 \sigma_{w,i}^2
   = \sum_i\biggl(\frac{\mathbf{1}_{x_i\in V}-\bar{I}}{\sum_{j}w_{j}}\biggr)^2\sigma_{w,i}^2\; .
  \label{eq:sigmamodel}
\end{align}
For a partition of the fiducial region into several bins $V_i$ both uncertainties add in quadrature,
\begin{align}
  \sigma_\text{stat}^2 = \sum_i\sigma_{\text{stat},V_i}^2
  \qquad \text{and} \qquad 
  \sigma_\text{model}^2 = \sum_i\sigma_{\text{model},V_i}^2 \; .
  \label{eq:quadsum}
\end{align}

%%%%%%%%%%%%%%%
\subsubsection*{Amplification}

Following Section~3 of Ref.~\cite{Bahl:2025ryd}, we evaluate the averaging condition in Eq.\eqref{eq:amplification_core} based on the surrogate estimate of the fiducial rate defined in Eq.\eqref{eq:itrue}
\begin{align}
    M_I\left[D_\text{true}^{\Ngen},\ptrue \right] \Bigg|_{\Ngen = \Nequiv} \really
    M_I\left[ D_\text{surr}^{\Ngen}, \ptrue \right] \Bigg|_{\Ngen \to \infty}
 \; .
\label{eq:def_mi}
\end{align}
For the right-hand size we define the measure
\begin{align}
  M_I\bigl[D_\text{surr}^{\Ngen},\ptrue\bigr]
   = \XLangle
      \bigl(I(\ptrue)-\bar I(D_\text{surr}^{\Ngen})\bigr)^2
  \XRangle 
  = \underbrace{\sigma_\text{stat}^2}_{\propto\,1/\Ngen}
  + \underbrace{\sigma_\text{model}^2}_{\text{const.\ in}\ \Ngen} .
  \label{eq:MI_decomp}
\end{align}
On the left-hand side of Eq.\eqref{eq:def_mi} only the statistical uncertainty contributes. Using Eq.\eqref{eq:sigmastat_neff}, the implicit definition of $\Nequiv$ in terms of $\Neff$ reads
\begin{align}
  \sigma_\text{stat}^2 \Bigg|_{\Neff = \Nequiv} = \sigma_\text{model}^2
  \qquad\Leftrightarrow\qquad
  \Nequiv = \frac{\bar I\, ( 1 - \bar I)}{\sigma_\text{model}^2} \; .
  \label{eq:nequiv}
\end{align}
%

%%%%%%%%%%%%%%%%%%%%%%%%%%%%%%
\subsection{Differential amplification factor}
\label{sec:diffamp}

Averaging amplification is local and does not capture features below the resolution defined by $\bar I$. For an alternative, differential amplification~\cite{Bahl:2025ryd} we turn Eq.\eqref{eq:amplification_core} into a comparison of two datasets,
\begin{align}
  M_I\bigl[D_\text{surr}^{\Ngen},\ptrue\bigr]
  \quad \to\quad 
  M\bigl[D_\text{surr}^{\Ngen},D_\text{true}^{\Nholdout}\bigr] \; ,
  \label{eq:onetwo}
\end{align}
where $D_\text{true}^{\Nholdout}$ is an independent truth dataset of size $\Nholdout$. For our weighted evaluation, $\Ngen$ will be phrased in terms of $\Neff$, as defined in Eq.\eqref{eq:neff}.

%%%%%%%%%%%%%%%
\subsubsection*{Kolmogorov--Smirnov test}

The differential amplification condition requires a two-sample test statistic for which the left-hand side of Eq.\eqref{eq:amplification_core} is known, \ie with a known asymptotic behavior for two samples drawn from the same distribution. This requirement singles out the Kolmogorov--Smirnov (KS) test, which is cheap to evaluate and has a known asymptotic behavior.

As a uni-variate test requires a summary statistic for discriminating two datasets. According to the Neyman--Pearson lemma the optimal test statistic is the (log-)likelihood ratio
\begin{align}
  T(x)
  &= \log\frac{p_\text{surr}(x)}{\ptrue(x)} \notag \\
  &= \log\frac{|\mathcal{M}|_\text{surr}^2(x)}{|\mathcal{M}|_\text{true}^2(x)} 
  = \log w(x) \; .
  \label{eq:logllr}
\end{align}
In contrast to the generative network case, where this ratio has to be approximated by a trained classifier, the reweighting in Eq.\eqref{eq:weight} provides us with the density ratio $w(x)$ explicitly.

The KS distance between two 1D datasets is defined in terms of their empirical cumulative distribution functions $F(t,D)$,
%Q
\begin{align}
  M_\text{KS}\bigl[D_{1},D_{2}\bigr] = \sup_{t}\bigl|F(t,D_{1})-F(t,D_{2})\bigr| 
  \qquad \text{with} \qquad 
  F(t,D)=\frac{1}{n}\sum_{x_i\in D}\mathbf{1}_{T(x_i)\le t}  \; .
  \label{eq:MKS}
\end{align}
If both samples are drawn from the same distribution, the rescaled KS statistic asymptotically follows the Kolmogorov distribution $p_{K}(K)$,
\begin{align}
  \sqrt{\frac{n_1 \,n_2}{n_1+n_2}}\,M_\text{KS}\bigl[D_1,D_2\bigr]
   = K\sim p_{K}(K) .
  \label{eq:Knull}
\end{align}
The scaling factor absorbs the two known statistical uncertainties $1/n_1+1/n_2$. As we are only interested in the finite expectation value of this rescaled statistic, we use the expectation value 
\begin{align}
  \sqrt{\frac{n_1\,n_2}{n_1+n_2}}\,\bigl\langle M_\text{KS}\bigr\rangle
   = \langle K\rangle = \sqrt{\frac{\pi}{2}}\,\log 2 \; .%\approx 0.869 .
  \label{eq:Kexp}
\end{align}
In terms of the KS distance, the implicit definition of $\Nequiv$ in Eq.\eqref{eq:amplification_core} becomes
\begin{align}
  M_\text{KS}\bigl[D_\text{true}^{\Ngen},D_\text{true}^{\Nholdout}\bigr] \Bigg|_{\Ngen = \Nequiv}
  \;\stackrel{!}{=}\;
  M_\text{KS}\bigl[D_\text{surr}^{\Ngen},D_\text{true}^{\Nholdout}\bigr] \Bigg|_{\Ngen \to \infty}  \; .
\end{align}
For the left-hand side, we use the above scaling, while for the right-hand side the independent truth reference keeps the unweighted form while the generated dataset is represented by weighted truth events,
\begin{align}
  F(t,D_\text{true}^{\Nholdout})
  &=\frac{1}{\Nholdout}\sum_{j=1}^{\Nholdout}\mathbf{1}_{T(x_{j})\le t}\;,  \notag \\
  F(t,D_\text{surr}^{\Ngen})
  &= \frac{1}{\sum_iw_i} \sum_j \,w_j \,\mathbf{1}_{T(x_j)\le t} \;.
  \label{eq:weightedcdf}
\end{align}
This allows us to evaluate
\begin{align}
  \langle K\rangle\,\sqrt{\frac{\Nequiv+\Nholdout}{\Nequiv\,\Nholdout}}
  \; &\stackrel{!}{=}\;
  \XLangle M_\text{KS}\bigl[D_\text{surr}^{\Ngen},D_\text{true}^{\Nholdout}\bigr] \XRangle \Bigg|_{\Ngen \to \infty} 
  \equiv M_\text{KS,bias} \notag\\
  \Rightarrow \qquad 
  \Nequiv
  &= \frac{\Nholdout}{\Nholdout\bigl(M_\text{KS,bias}/\langle K\rangle\bigr)^2-1} \; ,
  \label{eq:nequiv_KS}
\end{align}
with the convention $\Nequiv \to \infty$ when the denominator is non-positive and the bias lies below the noise floor. The weakness of the KS test is that the above condition becomes independent of $\Nequiv$ for $\Nequiv\gg \Nholdout$.

The KS statistic is distributed according to the rescaled Kolmogorov distribution, so we can quantify deviations using its confidence levels. With probability $1-\alpha$ the Kolmogorov variable lies in
\begin{align}
  \langle K\rangle - c(\alpha) < K < \langle K\rangle + c(\alpha)
  \qquad \text{with} \qquad 
  c(\alpha)\simeq\sqrt{\frac{1}{2}\log \frac{2}{\alpha} } \; ,
  \label{eq:KSconf}
\end{align}
so we can obtain a confidence interval on the KS-amplification factor $G^\text{KS}$ by replacing $\langle K\rangle$ with $\langle K\rangle\pm c(\alpha)$ in Eq.\eqref{eq:nequiv_KS}. The KS-metric can be localized in a fiducial region $V$ by only considering events that fall into this region.

%%%%%%%%%%%%%%%%%%%%%%%%%%%%%%
\subsection{Stratified estimation for kinematic tails}
\label{sec:stratified}

Both amplification measures require sufficiently many truth events. If $V$ lies in a kinematic tail with $\ptrue(x) \ll 1$, the small test sample makes the ratio-estimator variance and the KS noise floor large. A practical remedy is to generate $n_\text{targ}$ additional targeted points in a phase-space region $W \supset V$. They are drawn from
\begin{align}
  p_{W}(x) = \frac{1}{P_W} \ptrue(x)\,\mathbf{1}_{x\in W} 
  \qquad \text{with} \qquad 
  P_W = \int_W \text{d}x \ptrue(x) 
      = \frac{1}{n_\text{test}}
    \sum_{k=1}^{n_\text{test}}\mathbf{1}_{x_k\in W} \; ,
\end{align}
estimating the normalization from the truth test sample.

For averaging amplification, the ratio estimator applied to the targeted sample gives the conditional weighted fraction
\begin{align}
  \bar I_{V|W}
  = \frac{\sum_{i\in V} w_i}{\sum_{j\in W} w_j}
  \;\xrightarrow{n_\text{targ}\to\infty}\;
  \frac{\int_V \text{d}x \; w(x)\, \ptrue(x)\,}{\int_W \text{d}x \; w(x)\, \ptrue(x)},
\end{align}
where the sums run over the targeted events. The unconditional surrogate fraction is 
\begin{align}
    \bar I_V = P_W\cdot \bar I_{V|W} \; .
\end{align}
Linear propagation of the per-event weight uncertainties through the conditional ratio yields
\begin{align}
  \sigma_\text{model}^2
  = P_W^2\,\sigma_\text{model,cond}^2
  \qquad \text{and} \qquad 
  \sigma_\text{stat}^2
    = P_W^2\,\sigma_\text{stat,cond}^2
      + \bar I_{V|W}^2\,\text{Var}(P_W) \; ,
  \label{eq:strat_var}
\end{align}
where $\sigma_\text{model,cond}^2$ and $\sigma_\text{stat,cond}^2$ are the standard ratio-estimator uncertainties computed from the targeted sample alone, and $\text{Var}(P_W)$ is given by the binomial variance
\begin{align}
    \text{Var}(P_W)= \frac{P_W(1-P_W)}{n_\text{test}} \;.
\end{align}

For the differential KS estimator, the targeted sample replaces the original test sample as the weighted evaluation set for bins inside $W$. A second independent targeted truth dataset serves as the KS reference. 

\clearpage
%%%%%%%%%%%%%%%%%%%%%%%%%%%%%%%%%%%%%%%%%%%%%%%%%%%
\section{Gluon-associated \texorpdfstring{$Z$}{Z} production}
\label{sec:results}

We use the established production of an on-shell $Z$ boson together with a variable number of gluons~\cite{Badger:2022hwf},
\begin{align}
  q\bar q \;\to\; Z + n_g\,g \eqcomma
  \qquad n_g = 1,\dots,4 \eqperiod
\end{align}
We generate events with MadGraph5\_aMC\textsc{@NLO}~\cite{Alwall:2014hca} and apply the global cuts
\begin{align}
  p_{T} > 20~\gev
  \qqqquad 
  \Delta R_{gg} > 0.4
\end{align}
to regulate the soft and collinear divergences. The squared matrix element $|\mathcal{M}|^2$ is the regression target. It is invariant under permutations of identical final-state gluons. Increasing $n_g$ increases the phase-space dimensionality and the complexity of the amplitude, making this process a perfect testing ground for amplification.

%%%%%%%%%%%%%%%
\subsubsection*{Architecture and training}

We use the Lorentz-Equivariant Geometric Algebra Transformer (L-GATr)~\cite{Brehmer:2024yqw,Petitjean:2025zjf,Favaro:2026amj} as a standard LHC transformer architecture. It maps the set of external momenta $x=(p_1,\dots,p_n)$ to the log-squared matrix element, to tame its wide dynamic range,
\begin{align}
  A(x)=\log \left| \mathcal{M} \right|^2(x) \; .
  \label{eq:target}
\end{align}
The L-GATr represents each external particle as a token and embeds their 4-momenta as multivectors in the geometric Clifford algebra $\mathbb{G}_{1,3}$ over 4-dimensional Minkowski space. All internal operations, the equivariant linear layers, the geometric product, and the attention based on the $\mathbb{G}_{1,3}$ inner product, are equivariant under the connected Lorentz group $\text{SO}^+(1,3)$. This way, the surrogate respects the spacetime symmetry of the underlying amplitude by construction. 

%----------------------------------------------------------
\begin{figure}[b!]
  \centering
  \includegraphics[width=0.49\textwidth]{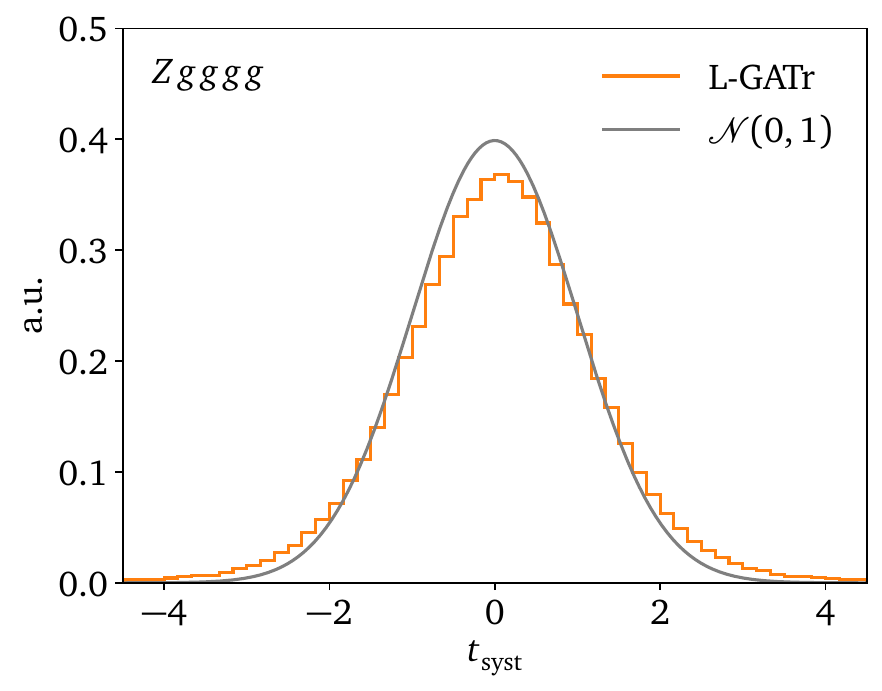}
\caption{Systematic pull for the $Z+4g$ surrogate trained on $10^5$ events.}
  \label{fig:pull-calibration}
\end{figure}
%----------------------------------------------------------

We use the optimized L-GATr-slim implementation~\cite{Petitjean:2025zjf,Favaro:2026amj} with a reduced operation set for fast training and evaluation. For the uncertainty-aware surrogate we replace the regression of Eq.\eqref{eq:target} by a heteroscedastic output consisting of a learned mean $A_\text{NN}(x)$ and uncertainty $\sigma_\text{syst}(x)$. Assuming a Gaussian likelihood leads to the heteroscedastic loss~\cite{Plehn:2022ftl,Badger:2022hwf,Bahl:2024gyt,Bahl:2025xvx},
\begin{align}
  \loss = %\left\langle
  \frac{\bigl[A_\text{NN}(x) - A_\text{true}(x)\bigr]^2}{2\,\sigma_\text{syst}^2(x)} + \log\sigma_\text{syst}(x) \; . %\right\rangle_{x\sim D_\text{train}}\; .
  \label{eq:het_loss}
\end{align}
The training amplitude points are generated and distributed as unit-weight events following the predicted kinematic distributions. At each point, the training either minimizes the numerator by learning the true amplitude or enlarges $\sigma_\text{syst}(x)$. Propagating $\sigma_\text{syst}(x)$ gives us a local uncertainty $\sigma_{|\mathcal{M}|^2}(x)$ needed for Eq.\eqref{eq:def_weights}. We give all hyperparameters in Appendix~\ref{app:hyperparams}.

We use one training dataset and two test datasets $\{ (A_\text{true}(x),x)\}$ with 1M points each. For each gluon multiplicity we train the amplitude surrogates on 10k, 100k, and 1M points. We test the surrogate through reweighting the first test dataset with the learned surrogate, replacing the usual generated data. The second test dataset serves as the reference in the corresponding histograms and to evaluate the amplification metrics. For each gluon multiplicity we repeat the same setup with two test datasets of 1M phase space points in a targeted tail region.

We first confirm that the learned uncertainty is calibrated, as studied for the same processes in Ref.~\cite{Bahl:2026qaf}. In Figure~\ref{fig:pull-calibration}, we show the systematic pull 
\begin{align}
  t_\text{syst}(x) \equiv \frac{A_\text{NN}(x) - A_\text{true}(x)}{\sigma_\text{syst}(x)}\;,
\end{align}
for the $Z+ 4g$ surrogate trained on $10^5$ events. It should follow a unit Gaussian for sufficiently many phase space dimensions~\cite{Bahl:2026qaf,Dubey:2026cts}. For our purposes, this calibration is sufficient.

%%%%%%%%%%%%%%%%%%%%%%%%%%%%%%%%%%%%%%%%%%%%%%%%%%%
\subsection{\texorpdfstring{$Z+g$}{Z+g} results}
\label{sec:results-zg}

%----------------------------------------------------------
\begin{figure}[b!]
  \includegraphics[width=0.49\textwidth]{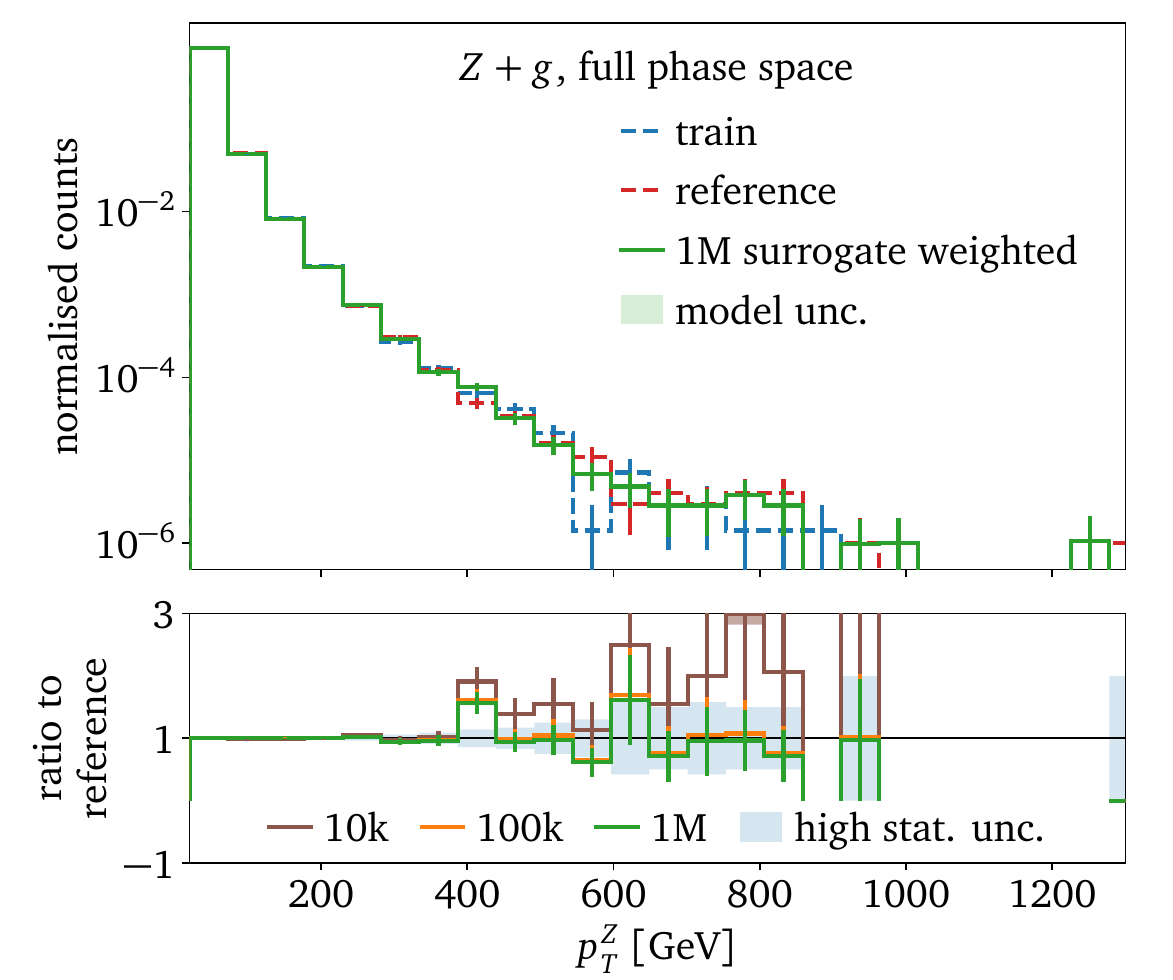}
  \includegraphics[width=0.49\textwidth]{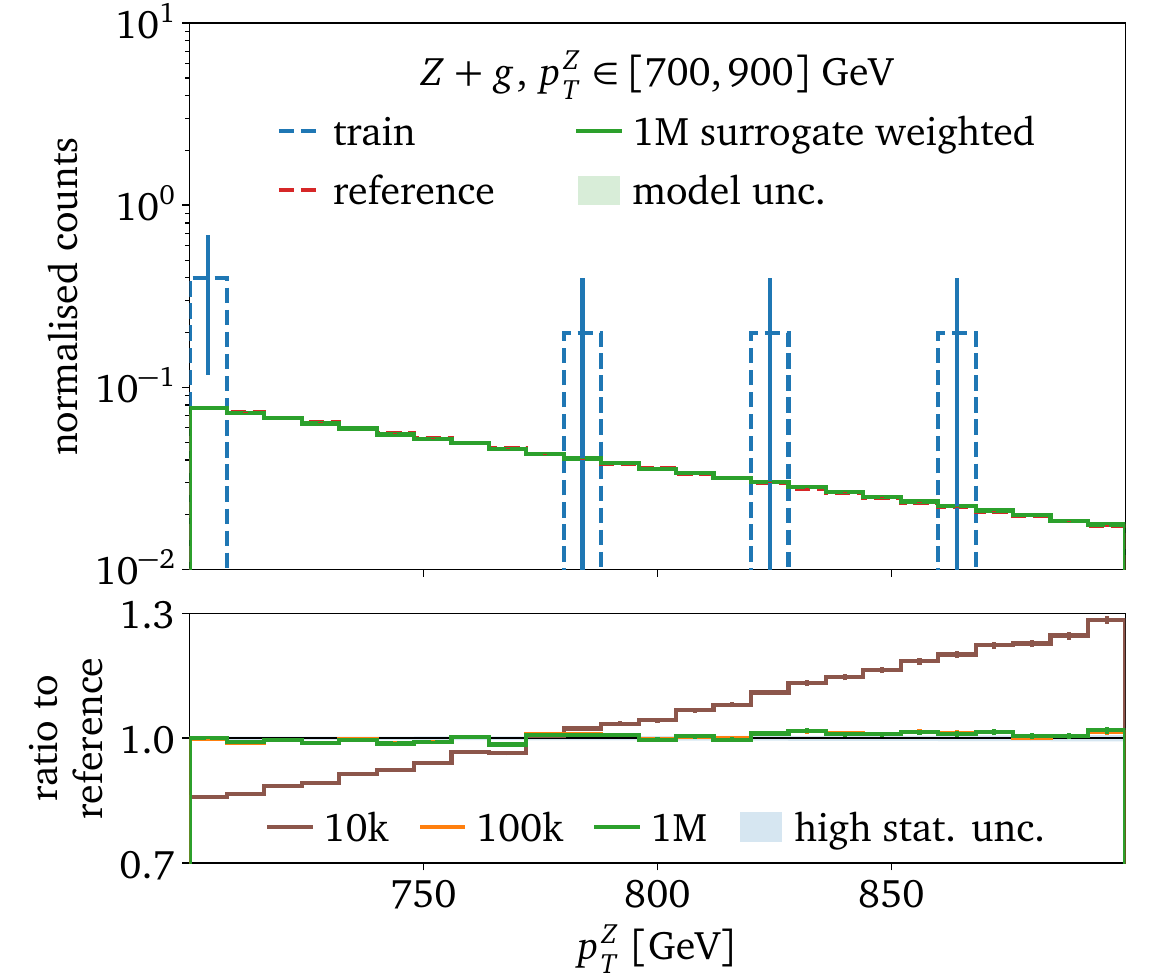}
\caption{Training dataset, reference sample, and weighted surrogate for the full $p_T^Z$ spectrum (left) and the tail region (right). The sub-panels show the ratios of the weighted surrogates (10k, 100k, and 1M training events) to the reference. Error bars denote the statistical uncertainty of the surrogate, the shaded band the statistical uncertainty of the reference.}
  \label{fig:zg_pt_diag_het}
\end{figure}
%----------------------------------------------------------

We first consider the simple process 
\begin{align}
 q \bar{q} \to Z+g 
\end{align}
We target the high-$p_T^Z$ region, specifically the window 
\begin{align}
 p_T^Z \in [700,800]~\gev 
\end{align}
for the stratified evaluation. 
%Inside the window we evaluate the local estimators on the targeted sample with 1M points in the tail window, outside the window on the original test sample with 1M points across phase space.

First, we validate the surrogate predictions against the reference sample over the full phase space and in the tail. In Figure~\ref{fig:zg_pt_diag_het} we show the $p_T^Z$ distributions and ratios to the reference sample for different training statistics. All ratios remain close to unity over the full $p_T^Z$ spectrum, at a somewhat reduced level for 10k training points and perfectly for 100k and 1M training points. This agreement carries on into the tail, with three training points in $p_T^Z \in [700,800]~\gev$ and two training points in $p_T^Z \in [800,900]~\gev$. Because of the high statistics of the targeted sample, the propagated model uncertainty is barely visible.

%----------------------------------------------------------
\begin{figure}[t]
  \includegraphics[width=0.49\textwidth]{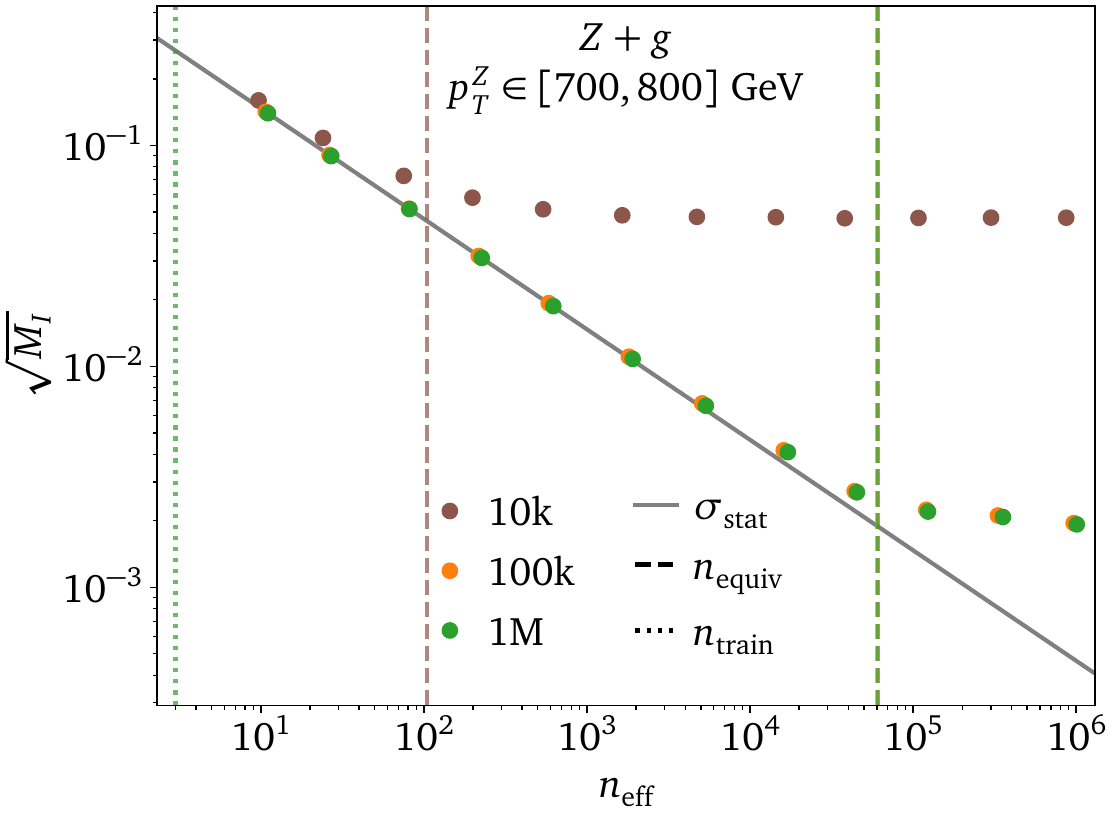}
  \includegraphics[width=0.49\textwidth]{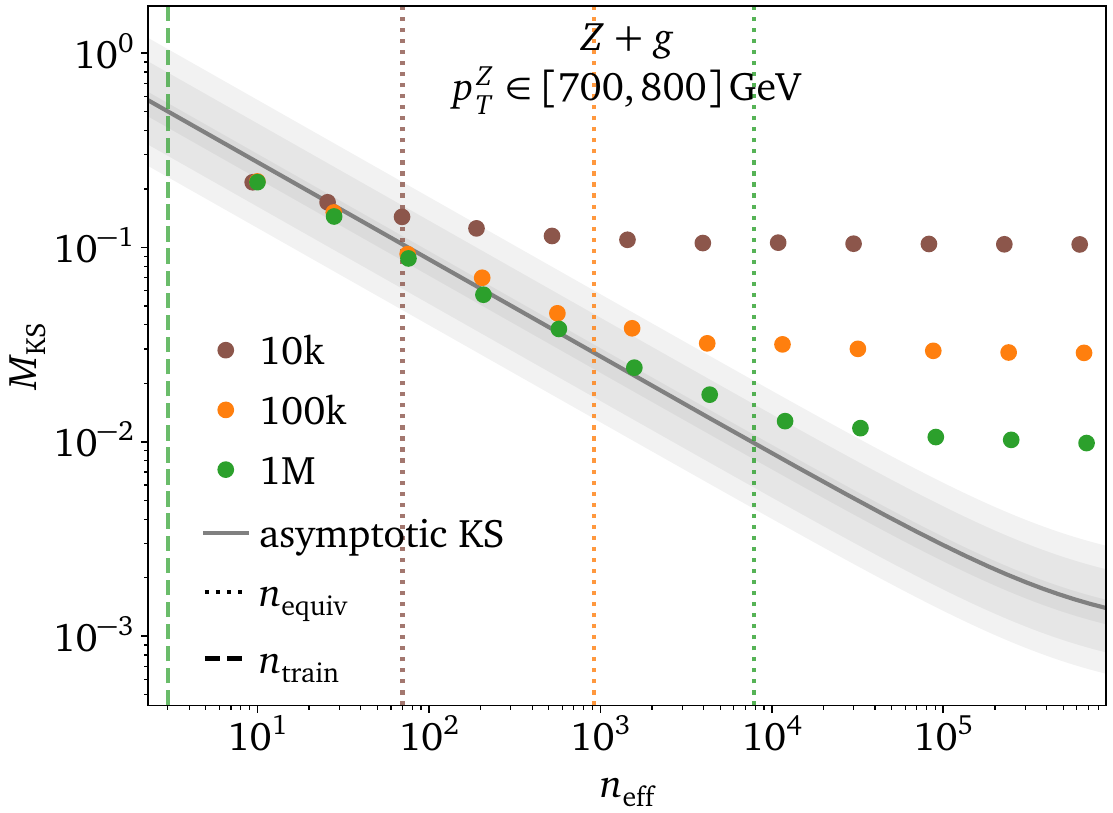}
\caption{Left: scaling of the averaging amplification metric with $\Neff$, including the expected statistical scaling. Right: scaling of the bootstrap-averaged differential amplification statistic, together with the theoretical expectation of Eq.\eqref{eq:Kexp}.}
  \label{fig:zg-700-800-scaling-bin1_het}
\end{figure}
%----------------------------------------------------------

After confirming the accuracy of our surrogate, we focus on the targeted tail region and study the scaling of the averaging and differential amplification metrics with $\Neff$. We implement the expectation value in Eq.\eqref{eq:MI_decomp} through bootstrap replicas of the weighted tail sample. We then compute $\bar I$ for each bootstrap replica and estimate the mean squared error
\begin{align}
M_I =
\frac{1}{N_\text{boot}}
\sum_{b=1}^{N_\text{boot}}
\left(\bar I_b-I_\text{true}\right)^2 \; .
\end{align}
Here, $I_\text{true}$ is evaluated on the sample. In the left panel of Figure~\ref{fig:zg-700-800-scaling-bin1_het} we show the scaling of $\sqrt{M_I}$ with $\Neff$ and compare it with the expected statistical scaling $M_I \propto 1/\Neff$. We indicate the three training points in this tail region for 1M training points, for 10k and 100k training points there are no events in this region. The three surrogates initially follow the expected statistical scaling and reach a plateau when the systematic model uncertainty limits the achievable precision. The surrogate trained on 10k events reaches the plateau considerably earlier, suggesting that 100k training points are sufficient to achieve the target accuracy for this simple process. The vertical dashed lines mark $\Nequiv$, corresponding to the point where the statistical uncertainty matches the systematic model uncertainty.

In the right panel of Figure~\ref{fig:zg-700-800-scaling-bin1_het} we show the scaling of the bootstrap-averaged KS statistic, with the theoretical expectation from Eq.\eqref{eq:Kexp}. We observe the deviation from the expected scaling for the different training datasets at increasing values of $\Neff \equiv \Nequiv$. The differential amplification is generally smaller than the averaging amplification, because it resolves deviations within the given bin~\cite{Bahl:2025ryd}.

For both amplification measures, the equivalent sample size relative to the number of training points gives us the amplification factor. Only the 1M event sample contains training points in the targeted region and gives us a finite amplification factor,
\begin{align}
 G = \frac{\Nequiv}{\Ntrain} \Bigg|_{[700,800]~\gev} = 
 \begin{cases}
 \dfrac{60.000}{3} = 20.000 \qquad & (M_I) \\[3mm]
 \dfrac{8.000}{3} = 2.670 \qquad & (M_\text{KS}) \; .
 \end{cases} 
\end{align}
%

%----------------------------------------------------------
\begin{figure}[t]
\centering
  \includegraphics[width=0.49\textwidth]{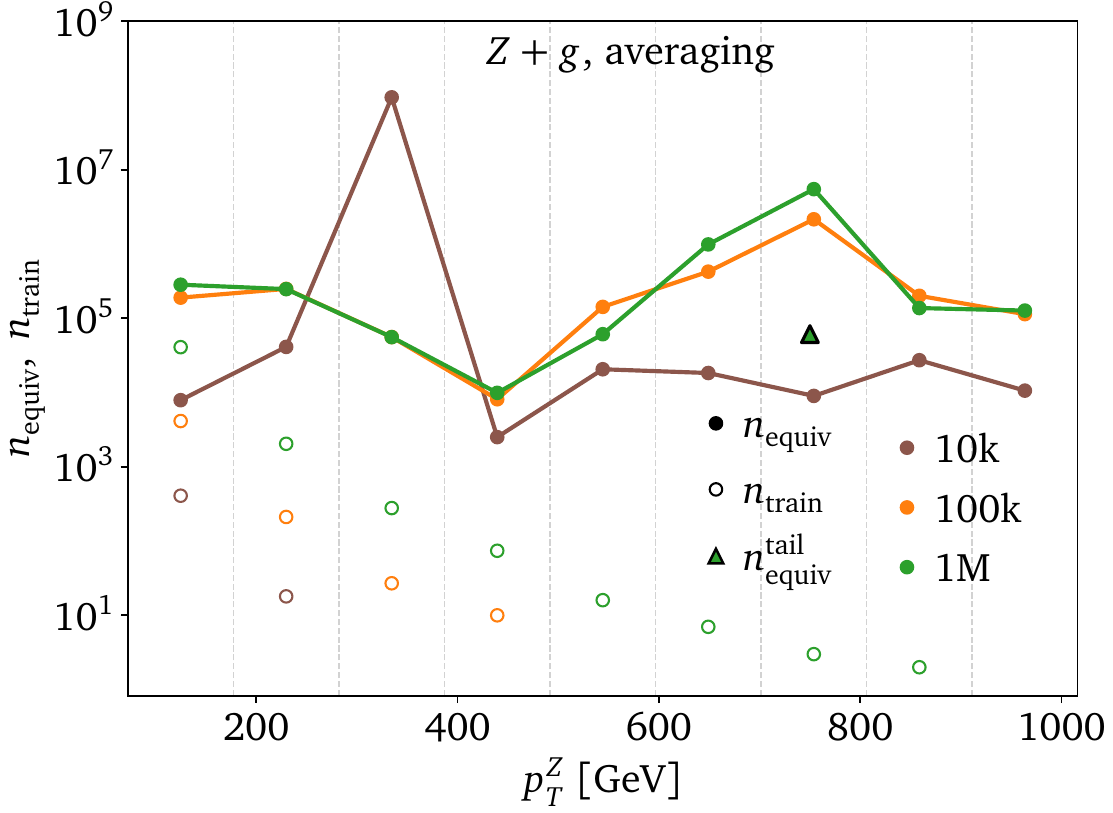}
\caption{Averaging amplification as a function of $p_T^Z$. For all bins, $\Nequiv$ is computed from the regular test dataset, for $p_T^Z \in [700,800]~\gev$ we also show the value computed from the targeted dataset with increased statistics.}
  \label{fig:zg-nequiv-het}
\end{figure}
%----------------------------------------------------------

In Figure~\ref{fig:zg-nequiv-het} we show how $\Nequiv$ from the averaging amplification varies across the $p_T^Z$ spectrum. We also show the corresponding training statistics $\Ntrain$. In contrast to Figure~\ref{fig:zg-700-800-scaling-bin1_het}, where we determine the amplification on the standard test dataset for each $p_T^Z$ bin. Across the $p_T^Z$ spectrum, we find $\Nequiv \gg \Ntrain$, albeit with strong fluctuations induced by the bin-wise evaluation. For 10k, 100k, and 1M training points the quality of the surrogate is roughly comparable and flat across $p_T^Z$. This suggests that, already for a limited number of training data points, only covering $p_T^Z \lesssim 200$~GeV reliably, the surrogate amplitude encodes the correct functional behavior. At the level of the central value it even extrapolates to $p_T^Z \gtrsim 1$~TeV. For completeness, we also show $\Nequiv$ calculated on the targeted sample for  $p_T^Z \in [700,800]~\gev$.

%%%%%%%%%%%%%%%%%%%%%%%%%%%%%%%%%%%%%%%%%%%%%%%%%%%
\subsection{\texorpdfstring{$Z+4g$}{Z+4g} results}
\label{sec:results-zgggg}

The next process we consider is at higher-multiplicity,
\begin{align}
    q \bar{q} \to Z+4g.
\end{align}
We present results for the intermediate-level $Z+2g$ and $Z+3g$ processes in Appendix~\ref{app:supplementary_results}. They exhibit the same qualitative behavior, so we focus on the two extreme cases. As before, we target the high-$p_T^Z$ tail
\begin{align}
 p_T^Z \in [1200,1400]~\gev \; .
\end{align}
Inside this window we evaluate the amplification measures on a targeted sample, outside on the original test sample.

\begin{figure}[t]
  \includegraphics[width=0.49\textwidth]{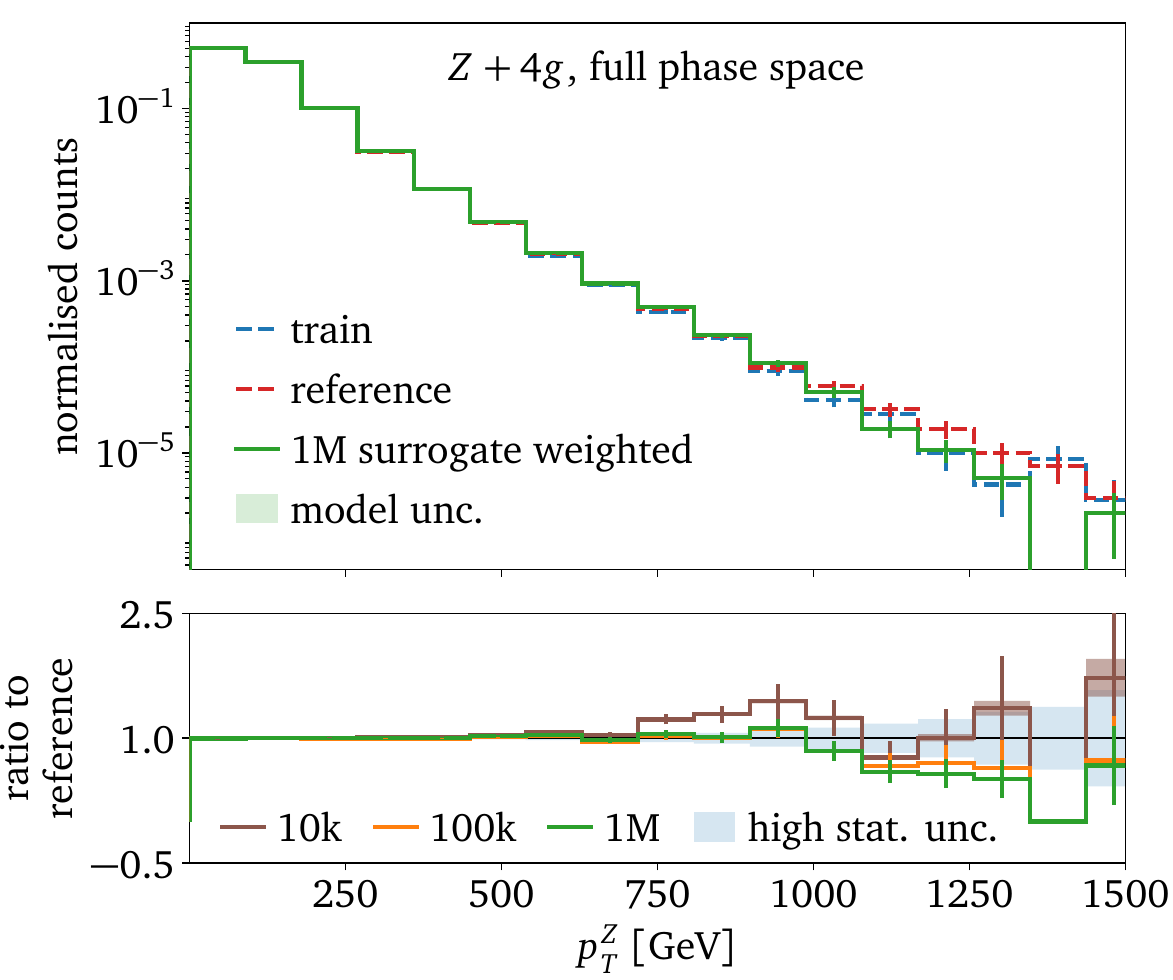}
  \includegraphics[width=0.49\textwidth]{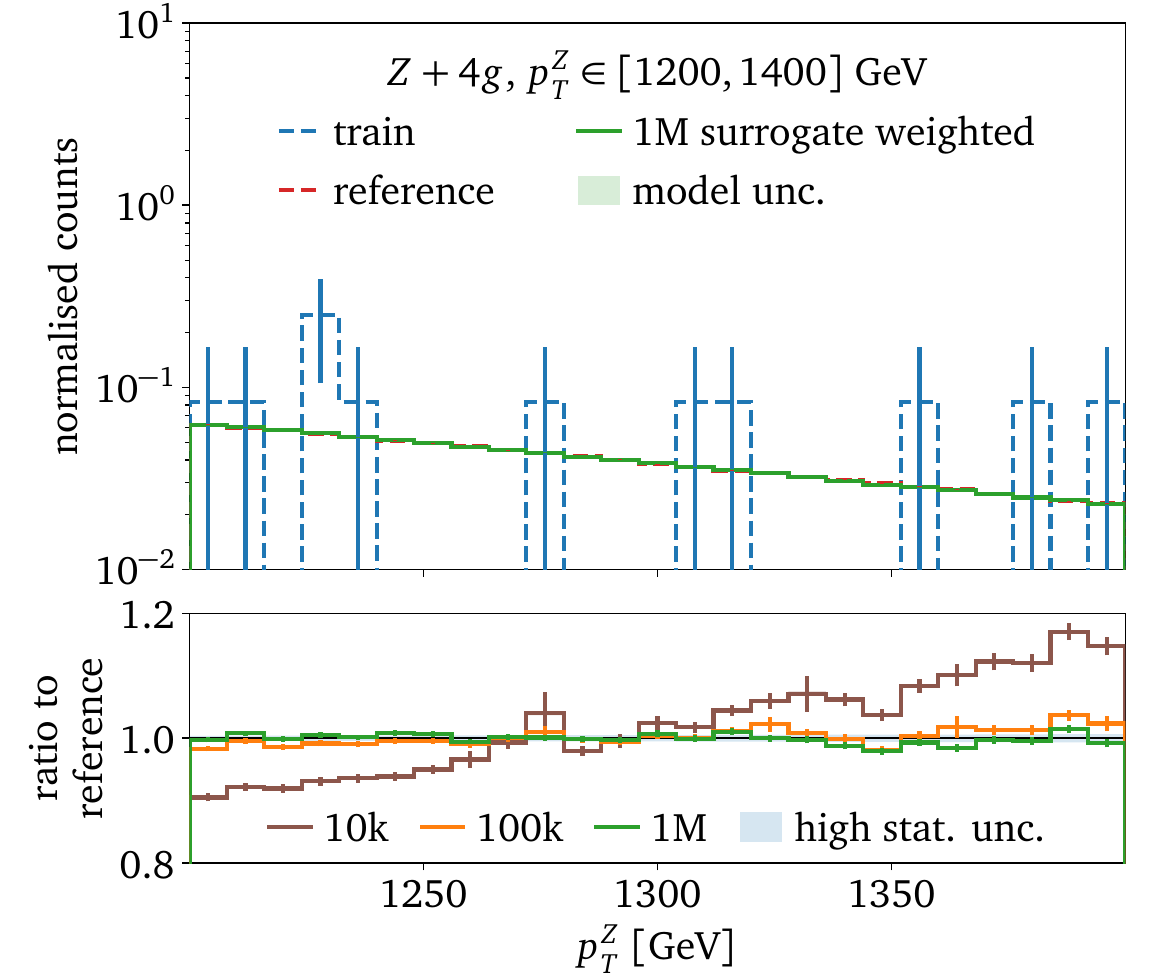}
\caption{Training dataset, reference sample, and weighted surrogate for the full $p_T^Z$ spectrum (left) and the tail region (right). The sub-panels show the ratios of the weighted surrogates (10k, 100k, and 1M training events) to the reference. Error bars denote the statistical uncertainty of the surrogate, the shaded band the statistical uncertainty of the reference.}
  \label{fig:zgggg_pt_diag_het}
\end{figure}

We first validate the surrogate predictions against the reference sample over the full phase space and in the tail region. Figure ~\ref{fig:zgggg_pt_diag_het} shows the $p_T^Z$ distributions and the ratios to the reference for three training sample sizes. Despite the increased complexity of the $Z+4g$ process, all three surrogates remain in good agreement with the reference sample, including the targeted region. The propagated model uncertainty is barely visible for the large targeted sample. We achieve this level of agreement even for only seven training points in $p_T^Z \in [1200,1300]~\gev$ and five points in $p_T^Z \in [1300,1400]~\gev$.

%----------------------------------------------------------
%----------------------------------------------------------

%----------------------------------------------------------
\begin{figure}[b!]
  \includegraphics[width=0.49\textwidth]{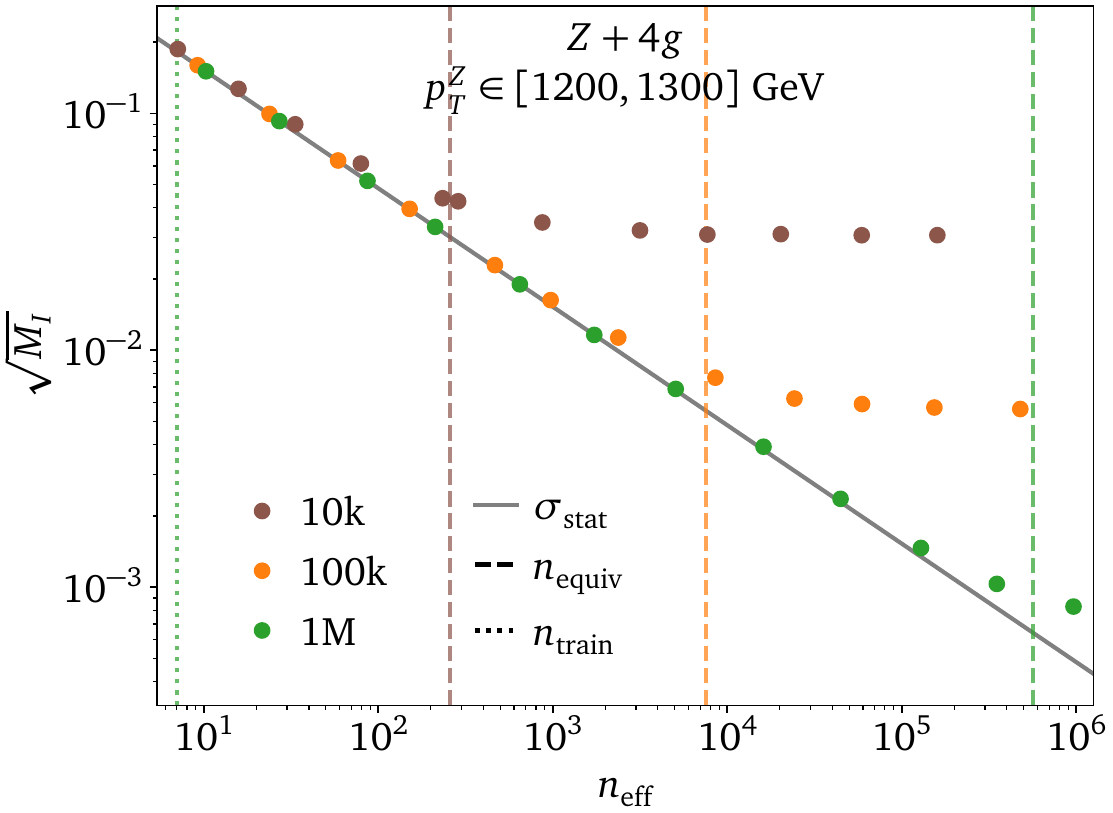}
  \includegraphics[width=0.49\textwidth]{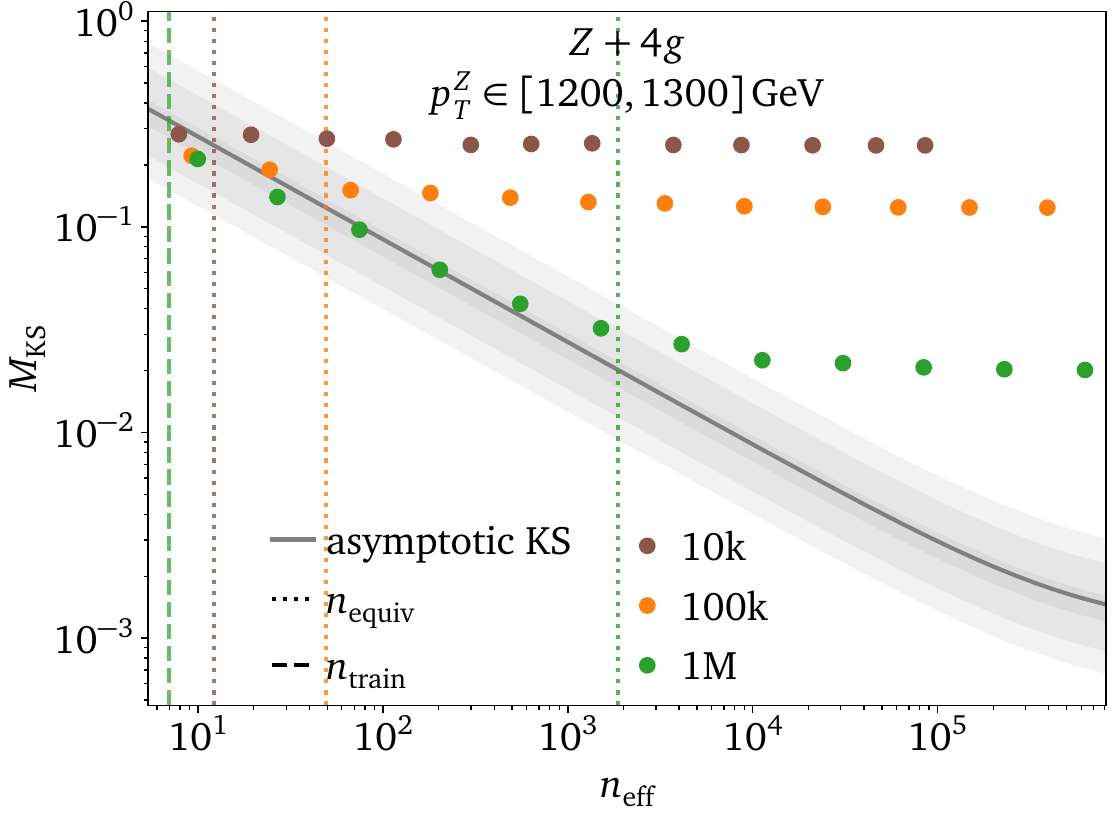}
\caption{Left: scaling of the averaging amplification metric with $\Neff$, including the expected statistical scaling. Right: scaling of the bootstrap-averaged differential amplification statistic, together with the theoretical expectation of Eq.\eqref{eq:Kexp}.}
  \label{fig:zgg-1200-1300-scaling-bin1_het}
\end{figure}
%----------------------------------------------------------

As before, we focus on the amplification in the kinematic tail region. In the left panel of Figure~\ref{fig:zgg-1200-1300-scaling-bin1_het} we again see that only the 1M training sample contains events in this region. Compared to the $Z + g $ process, the 100k and 1M training now reach the plateau at different $\Neff$ values, because the increased training statistics continues to improve the surrogate accuracy. 

In the right panel of Figure~\ref{fig:zgg-1200-1300-scaling-bin1_het} we show the scaling of the bootstrap-averaged KS statistic together with the expected statistical scaling. As for the averaging metric, the three surrogates depart from the statistical regime at progressively larger $\Nequiv$ values. Throughout, the differential amplification is smaller than the averaging amplification.

\begin{figure}[t]
  \centering
  \includegraphics[width=0.49\textwidth]{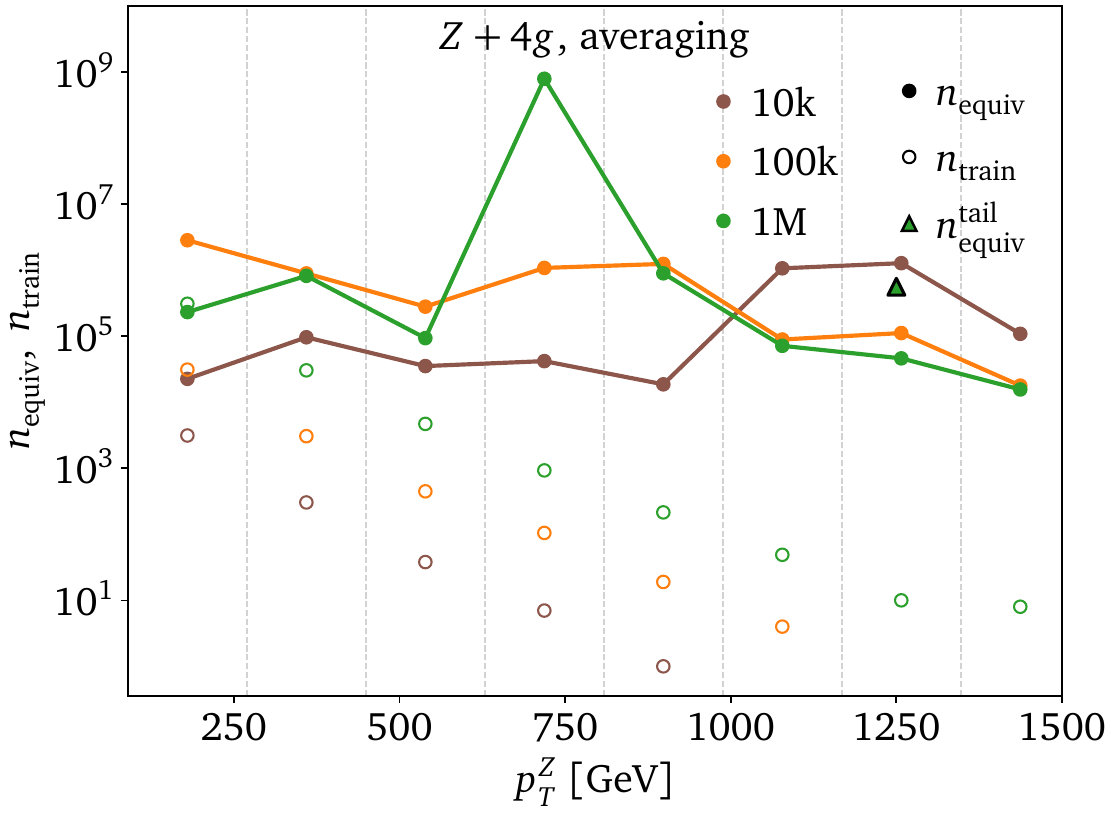}
\caption{Averaging amplification as a function of $p_T^Z$. For all bins, $\Nequiv$ is computed from the regular test dataset, for $p_T^Z \in [1200,1300]~\gev$ we also show the value computed from the targeted dataset with increased statistics.}
  \label{fig:zgggg-nequiv-het}
\end{figure}

We extract the amplification factor from the ratio of the equivalent sample size to the training statistics. Only the 1M training sample leads to finite ratios, 
\begin{align}
 G = \frac{\Nequiv}{\Ntrain} \Bigg|_{[1200,1400]~\gev} = 
 \begin{cases}
 \dfrac{560.000}{7} = 80.000 \qquad & (M_I) \\[3mm]
 \dfrac{2000}{7} = 290 \qquad & (M_\text{KS}) \; .
 \end{cases} 
\end{align}
For the averaging amplification we show in Figure~\ref{fig:zgggg-nequiv-het} the dependence of $\Nequiv$ and $\Ntrain$ on $p_T^Z$. As for the $Z+g$ process, we find $\Nequiv \gg \Ntrain$ across the spectrum. Moreover, $\Nequiv$ again remains fairly flat across $p_T^Z$. The sizeable bin-to-bin fluctuations arise from the statistical fluctuations of the individual bin fractions rather than from physically meaningful variations.

%----------------------------------------------------------
%----------------------------------------------------------

%%%%%%%%%%%%%%%%%%%%%%%%%%%%%%%%%%%%%%%%%%%%%%%%%%%
\section{Outlook}
\label{sec:outlook}

Precision simulations are a bottleneck of many LHC analyses, and network surrogates for loop amplitudes will help us overcome it. These surrogates leave the simulation chain structurally untouched and only replace a slow step. So far, they have been investigated in terms of $(i)$ accuracy, $(ii)$ calibrated uncertainties, and $(iii)$ numerical acceleration. In this paper, we have analyzed their statistical benefit: Can a surrogate trained on a limited number of exact amplitudes describe the underlying amplitude better than the training data does~\cite{Butter:2020qhk,Bahl:2025ryd}.

As example processes we chose $Z+ (1~...~4) g$ production. Since the density ratio to the truth is known per event, we avoided generating surrogate data and instead reweighted a truth sample. As statistical metrics, we adapted the averaging amplification, robust but local, and the differential amplification, optimal and global~\cite{Bahl:2025ryd}. Using a Lorentz-equivariant transformer, we found a significant amplification for all multiplicities considered. This amplification grows towards the kinematic tails, so the surrogate is worth the most precisely where the truth simulation runs out of events.

The comparison with generative networks turns this into a statement about the simulation chain. Using the same advanced transformer architecture, the amplitude surrogate amplifies much more than a generative network learning the full phase-space density~\cite{Bahl:2025ryd}. The reason is that a generative network has to learn the amplitude, the phase-space measure, the parton densities, and the cuts at once, and its amplification is limited by the hardest task. The surrogate only has to learn one smooth scalar field, which is genuinely expensive. 

%%%%%%%%%%%%%%%%%%%%%%%%%%%%%%%%%%%%%%%%%%%%%%%%%%%
\subsection*{Acknowledgements}

TP is supported by the Deutsche Forschungsgemeinschaft (DFG, German Research Foundation) under grant 396021762 -- TRR~257 \textsl{Particle Physics Phenomenology after the Higgs Discovery}. The authors acknowledge support by the state of Baden-Württemberg through bwHPC and the German Research Foundation (DFG) through grant no INST 39/963-1 FUGG (bwForCluster NEMO). RR acknowledges support by the IMPRS for Precision Tests of Fundamental Symmetries.

\appendix
%%%%%%%%%%%%%%%%%%%%%%%%%%%%%%%%%%%%%%%%%%%%%%%%%%%
\section{Network and training hyperparameters}
\label{app:hyperparams}

Table~\ref{tab:hyperparams} collects the L-GATr-Slim architecture and the training setup used for all $Z+n_g$ surrogates. The regression head outputs the pair $(\bar A, \log\sigma_\text{syst}^2)$, i.e.\ the mean of $\log|\mathcal{M}|^2$ and the per-event uncertainty entering the reweighting of Section~\ref{sec:reweight}. Training proceeds in two stages: a pure-MSE warm-up on the mean conditions the network before the heteroscedastic loss of Eq.\eqref{eq:het_loss} is switched on. Training, validation, and test events are drawn from statistically independent samples.

%----------------------------------------------------------
\begin{table}[h]
  \centering
  \begin{small}
  \begin{tabular}{ll}
    \hline
    \multicolumn{2}{l}{\textbf{L-GATr-Slim}} \\
    \hline
    Input channels & 1 multivector (four-momentum) + 3 scalar (particle type) \\
    Output channels & 2 scalar ($\bar A$, $\log \sigma_\text{syst}^2$) \\
    Hidden multivector channels & 40 \\
    Hidden scalar channels & 72 \\
    Attention blocks & 8 \\
    Attention heads & 8 \\
    MLP channels & 160 multivector / 288 scalar \\
    Attention channels per head & 5 multivector / 9 scalar \\
    \hline
    \multicolumn{2}{l}{\textbf{Training}} \\
    \hline
    Iterations & $200{,}000$ \\
    Batch size & $1024$ \\
    MSE warm-up iterations & $30{,}000$ ($10^4$, $10^5$ train.\ events); $100{,}000$ ($10^6$ train.\ events) \\
    Optimizer & Adam, $\beta = (0.99, 0.999)$, $\epsilon = 10^{-8}$, no weight decay \\
    Learning rate & $10^{-3}$ \\
    Scheduler & ReduceLROnPlateau (factor $0.3$, patience $10$) \\
    Validation frequency & every $1000$ iterations \\
    Early stopping & best-validation checkpoint restored \\
    Gradient-norm clipping & $5$ \\
    \hline
  \end{tabular}
  \end{small}
\caption{Hyperparameters of the L-GATr-Slim amplitude surrogate. Inputs are standardized and the gluon tokens are symmetrized to enforce permutation invariance. The scheduler patience is counted in validation steps, i.e.\ $10{,}000$ iterations.}
  \label{tab:hyperparams}
\end{table}
%----------------------------------------------------------

%%%%%%%%%%%%%%%%%%%%%%%%%%%%%%%%%%%%%%%%%%%%%%%%%%%
\section{Supplementary results}
\label{app:supplementary_results}
In this Appendix, we collect the results for the two intermediate processes, $Z+2g$ and $Z+3g$. They exhibit the same qualitative behavior as the $Z+g$ and $Z+4g$ processes discussed in the main text.

%%%%%%%%%%%%%%%%%%%%%%%%%%%%%%%%%%%%%%%%%%%%%%%%%%%
\subsection{\texorpdfstring{$Z+2g$}{Z+2g} results}
\label{sec:results-zgg}
\begin{figure}[H]
  \includegraphics[width=0.49\textwidth]{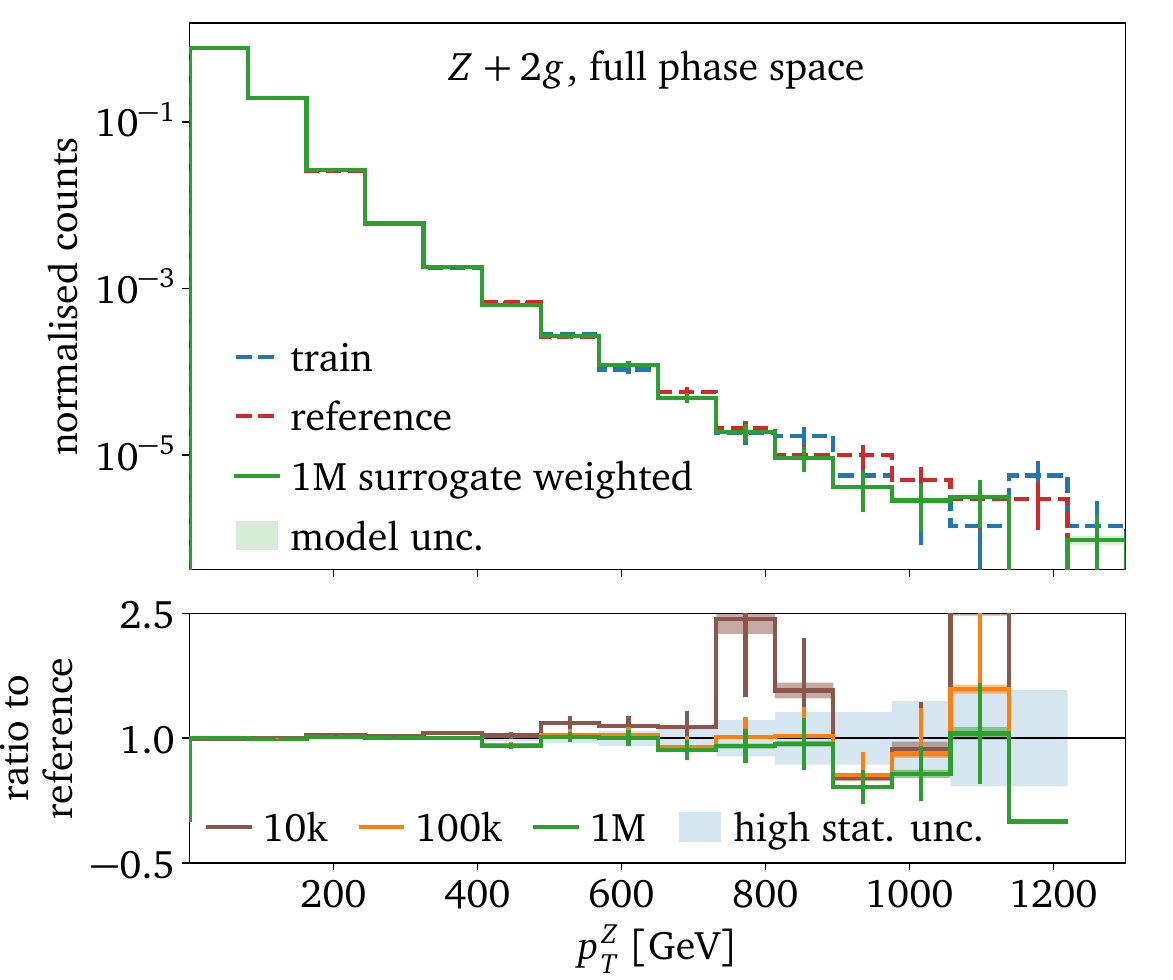}
  \includegraphics[width=0.49\textwidth]{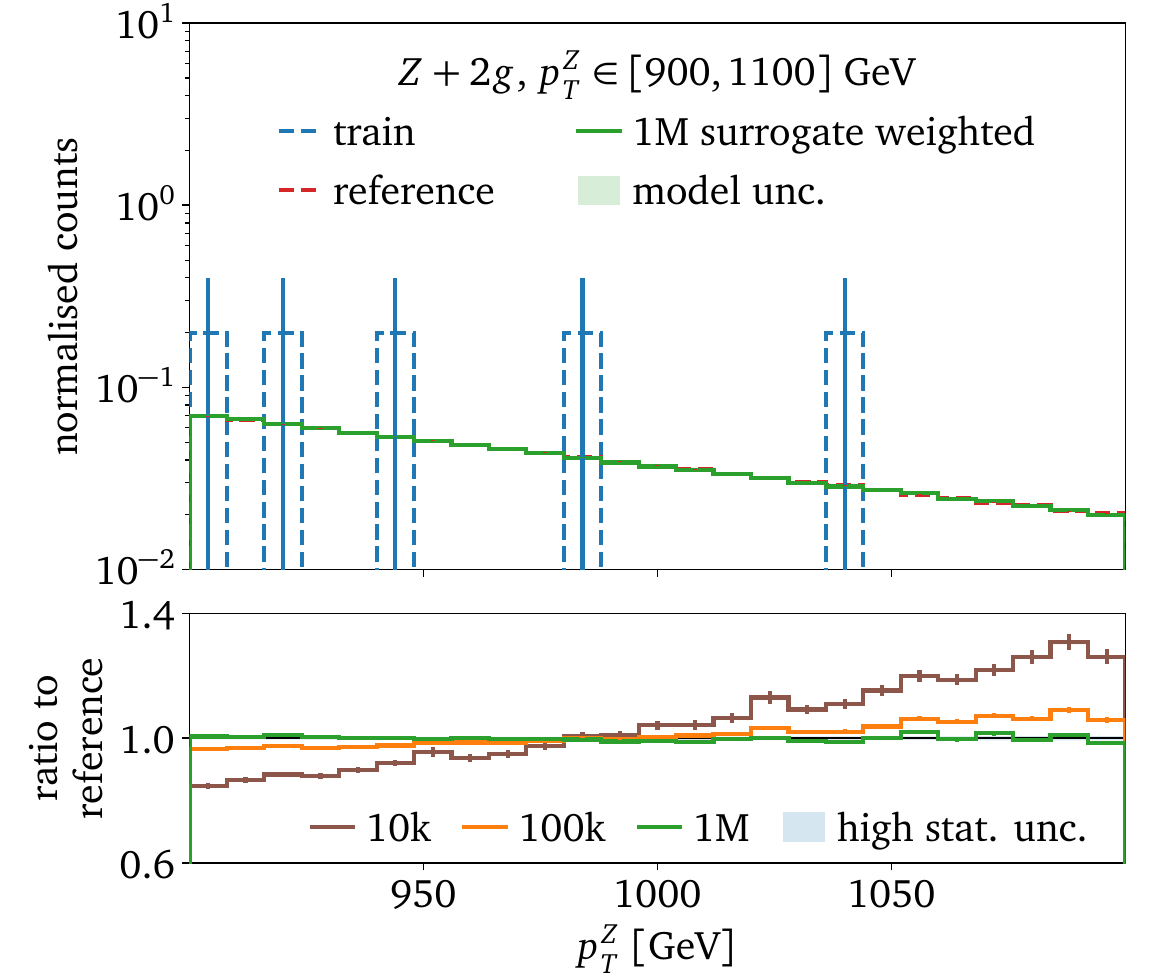}
\caption{Training dataset, reference sample, and weighted surrogate for the full $p_T^Z$ spectrum (left) and the tail region (right). The sub-panels show the ratios of the weighted surrogates (10k, 100k, and 1M training events) to the reference. Error bars denote the statistical uncertainty of the surrogate, the shaded band the statistical uncertainty of the reference.}
  \label{fig:zgg_pt_diag_het}
\end{figure}

We first consider the process $q \bar{q} \to Z+2g.$ The weighted surrogate predictions remain in excellent agreement with the reference sample over the full phase space and in the targeted tail region, as we show in Figure~\ref{fig:zgg_pt_diag_het}.

%----------------------------------------------------------
\begin{figure}[H]
  \includegraphics[width=0.49\textwidth]{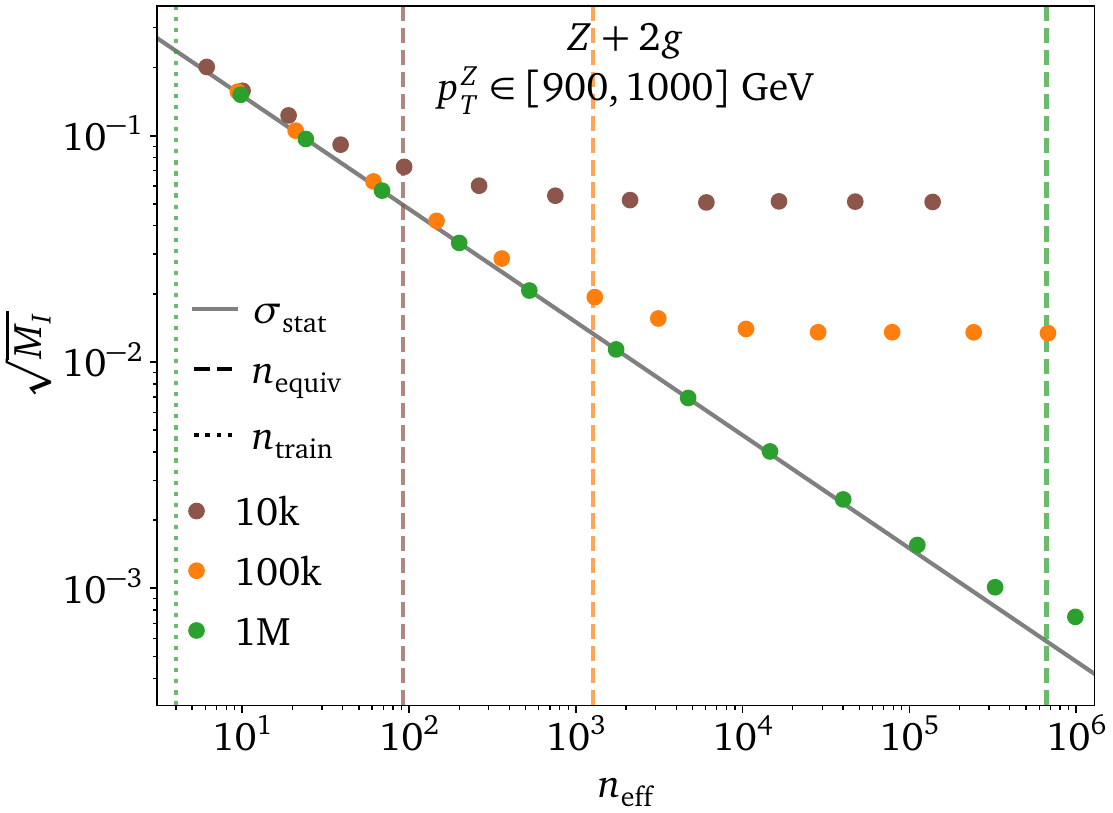}
  \includegraphics[width=0.49\textwidth]{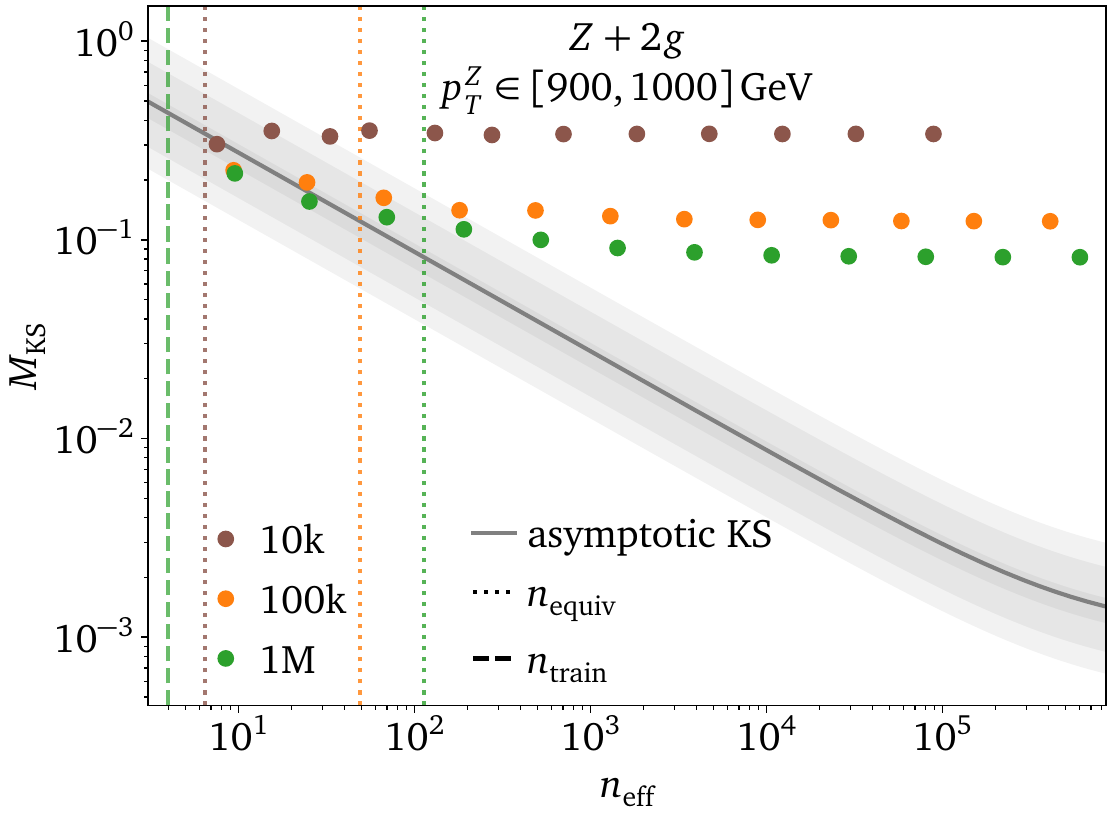}
\caption{Left: scaling of the averaging amplification metric with $\Neff$, including the expected statistical scaling. Right: scaling of the bootstrap-averaged differential amplification statistic, together with the theoretical expectation of Eq.\eqref{eq:Kexp}.}
  \label{fig:zgg-900-1000-scaling-bin1_het}
\end{figure}

%----------------------------------------------------------
Figure~\ref{fig:zgg-900-1000-scaling-bin1_het} shows that both the averaging and differential amplification metrics follow the expected statistical scaling before reaching the plateau, where the systematic model uncertainty becomes dominant. The corresponding equivalent sample sizes yield the amplification factor
\begin{align}
 G = \frac{\Nequiv}{\Ntrain} \Bigg|_{[900,1000]~\gev} = 
 \begin{cases}
 \dfrac{660.000}{4} = 165.000 \qquad & (M_I) \\[3mm]
 \dfrac{110}{4} = 28 \qquad & (M_\text{KS}) \; .
 \end{cases} 
\end{align}
%

%----------------------------------------------------------
\begin{figure}[H]
  \centering
  \includegraphics[width=0.49\textwidth]{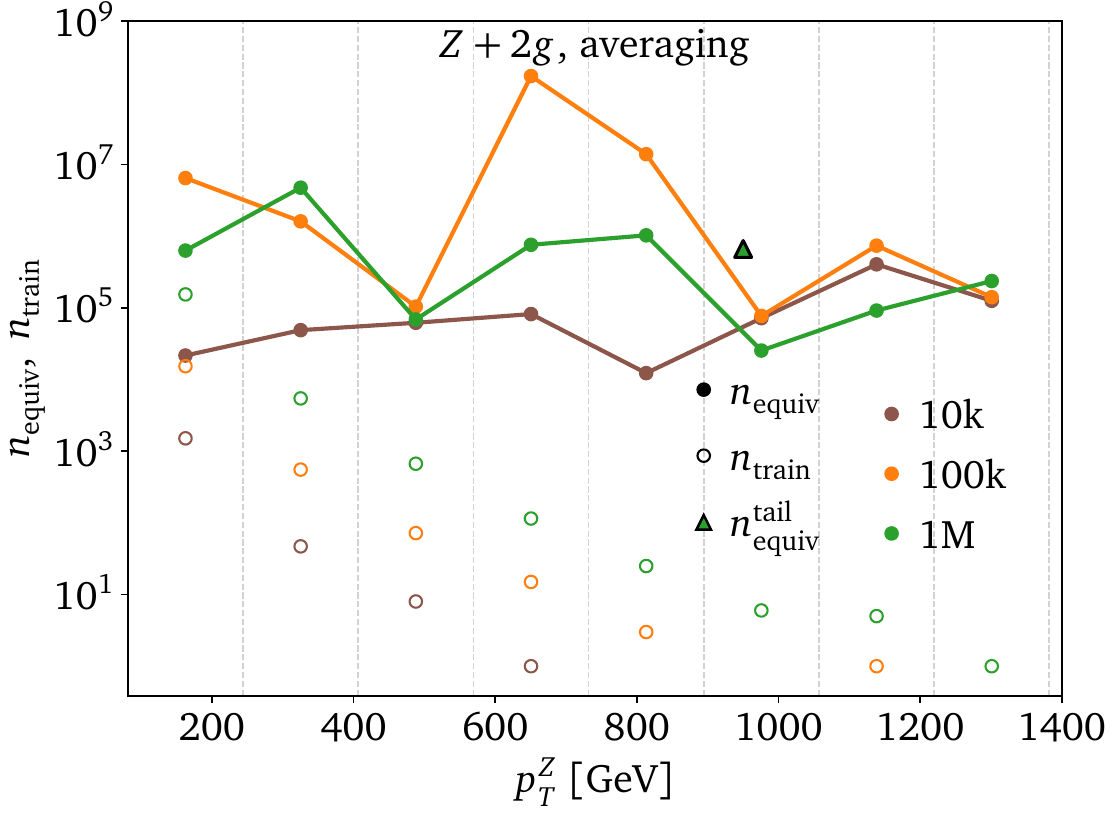}
\caption{Averaging amplification as a function of $p_T^Z$. For all bins, $\Nequiv$ is computed from the regular test dataset, for $p_T^Z \in [900,1000]~\gev$ we also show the value computed from the targeted dataset with increased statistics.}
  \label{fig:zgg-nequiv-het}
\end{figure}

%----------------------------------------------------------
Finally, in Figure~\ref{fig:zgg-nequiv-het} we show that the equivalent sample size exceeds the available training statistics over most of the $p_T^Z$ spectrum, demonstrating significant statistical amplification for this intermediate process as well.
%----------------------------------------------------------
%----------------------------------------------------------
%%%%%%%%%%%%%%%%%%%%%%%%%%%%%%%%%%%%%%%%%%%%%%%%%%%
\subsection{\texorpdfstring{$Z+3g$}{Z+3g} results}
\label{sec:results-zggg}
\begin{figure}[H]
  \includegraphics[width=0.49\textwidth]{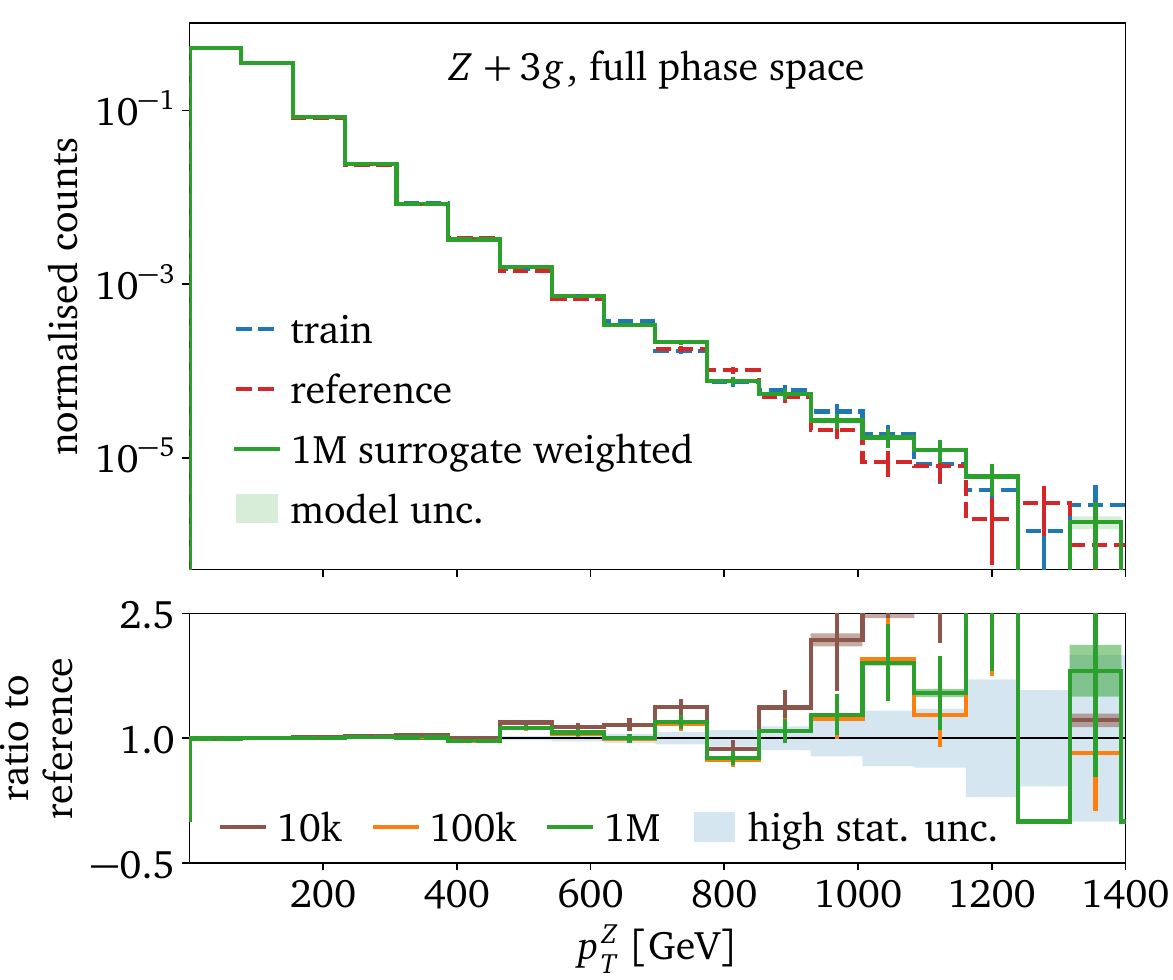}
  \includegraphics[width=0.49\textwidth]{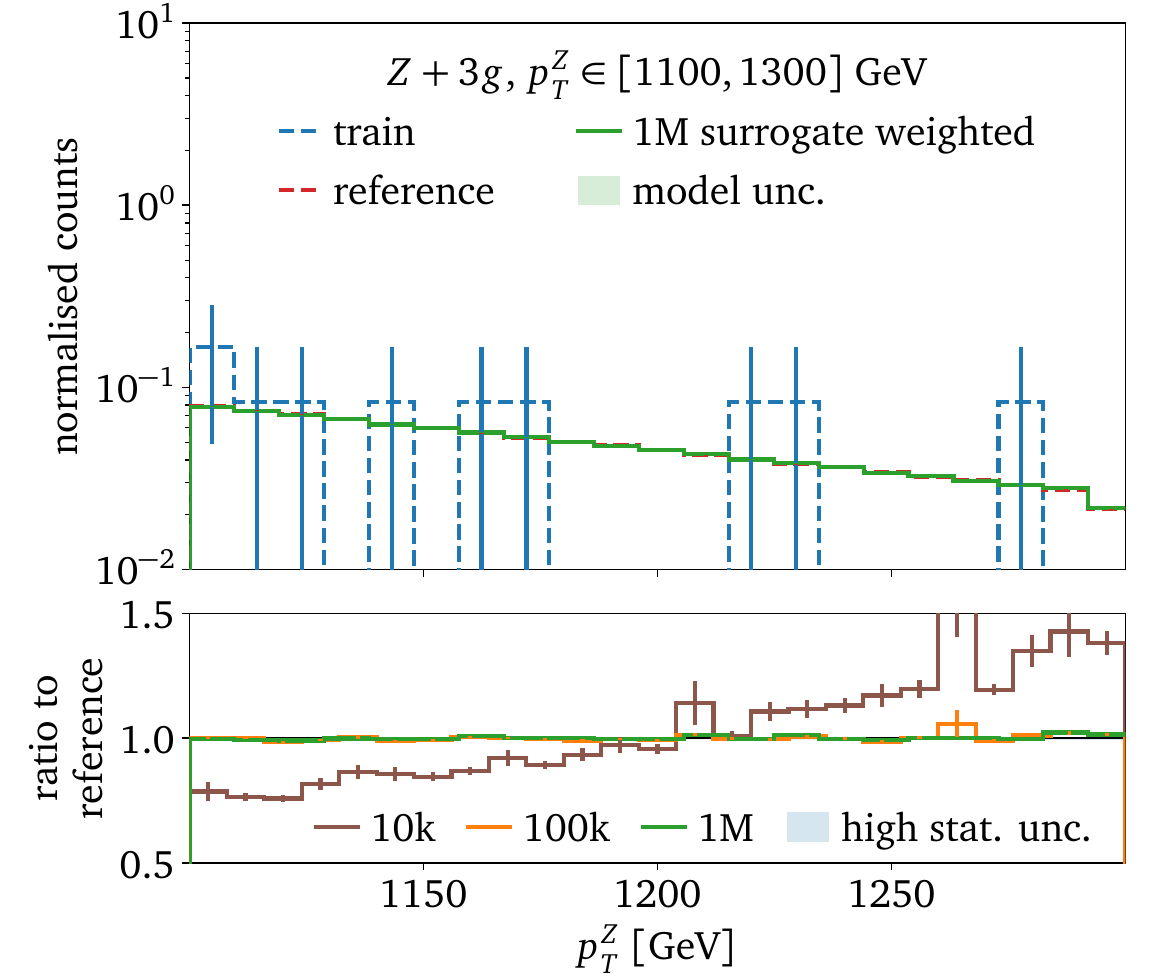}
\caption{Training dataset, reference sample, and weighted surrogate for the full $p_T^Z$ spectrum (left) and the tail region (right). The sub-panels show the ratios of the weighted surrogates (10k, 100k, and 1M training events) to the reference. Error bars denote the statistical uncertainty of the surrogate, the shaded band the statistical uncertainty of the reference.}
  \label{fig:zggg_pt_diag_het}
\end{figure}

As a final process, we consider $ q \bar{q} \to Z+3g$. The surrogate continues to accurately reproduce the reference sample over the full phase space as well as in the targeted tail region, as shown in Figure~\ref{fig:zggg_pt_diag_het}.

%----------------------------------------------------------
\begin{figure}[H]
  \includegraphics[width=0.49\textwidth]{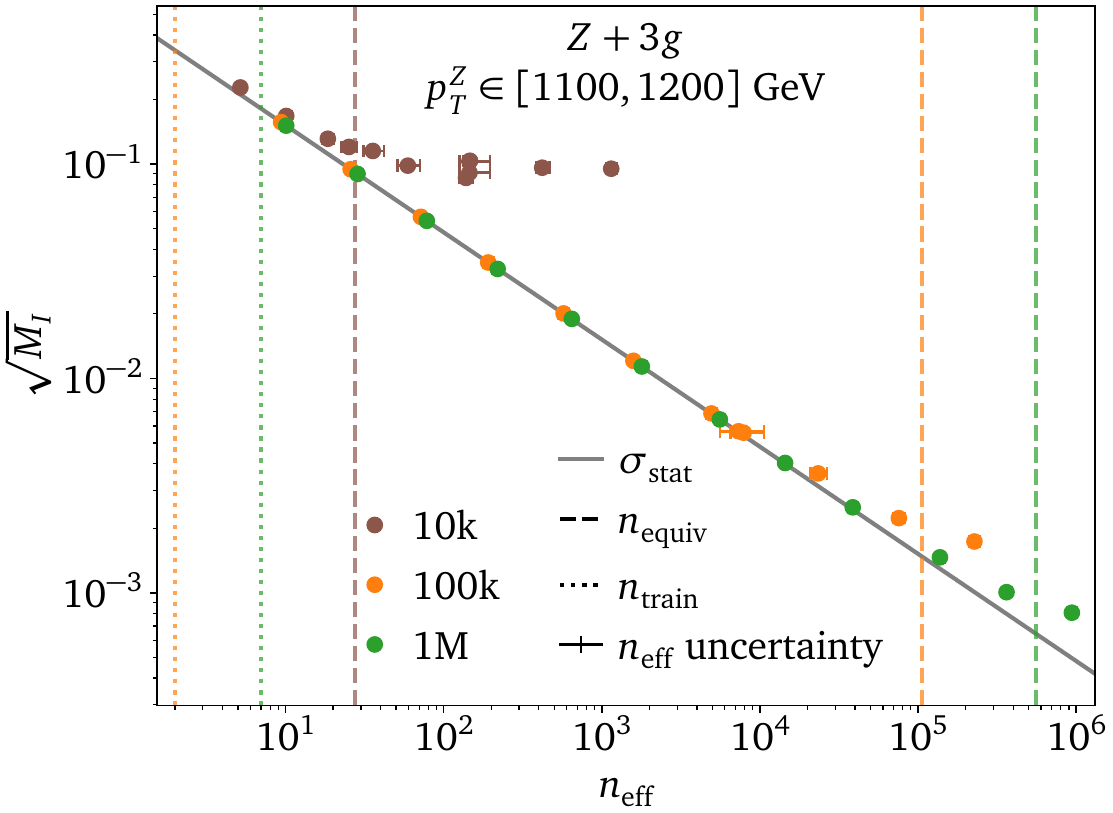}
  \includegraphics[width=0.49\textwidth]{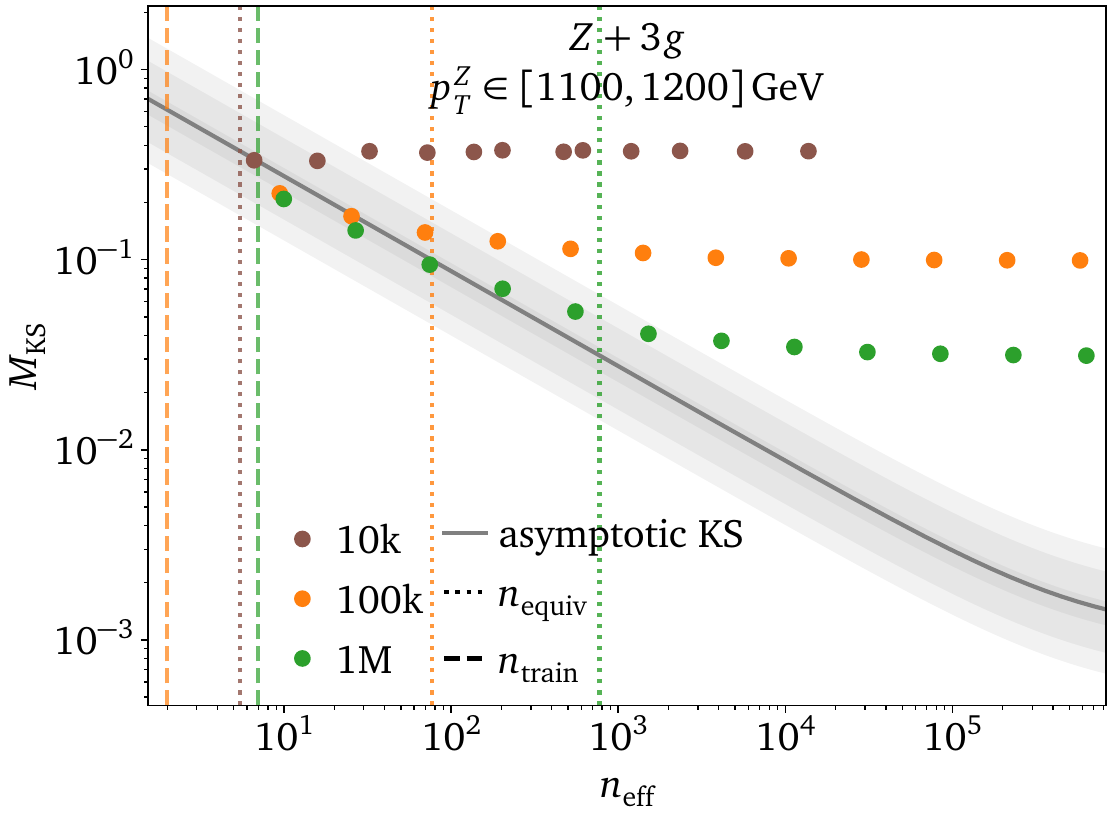}
\caption{Left: scaling of the averaging amplification metric with $\Neff$, including the expected statistical scaling. Right: scaling of the bootstrap-averaged differential amplification statistic, together with the theoretical expectation of Eq.\eqref{eq:Kexp}.}
  \label{fig:zgg-1100-1200-scaling-bin1_het}
\end{figure}

%----------------------------------------------------------
Figure~\ref{fig:zgg-1100-1200-scaling-bin1_het} shows that both amplification metrics follow the expected statistical scaling before reaching the plateau, where the systematic model uncertainty becomes the dominant contribution. For the averaging metric, we additionally show the bootstrap uncertainty on $\Neff$, indicating that the almost overlapping points of the 10k surrogate are consistent with the uncertainty on the effective sample size. The corresponding equivalent sample sizes yield the amplification factor
\begin{align}
 G = \frac{\Nequiv}{\Ntrain} \Bigg|_{[1100,1200]~\gev} = 
 \begin{cases}
 \dfrac{550.000}{7} = 79.000 \qquad & (M_I) \\[3mm]
 \dfrac{770}{110} = 290 \qquad & (M_\text{KS}) \; .
 \end{cases} 
\end{align}
%

%----------------------------------------------------------
\begin{figure}[H]
  \centering
  \includegraphics[width=0.49\textwidth]{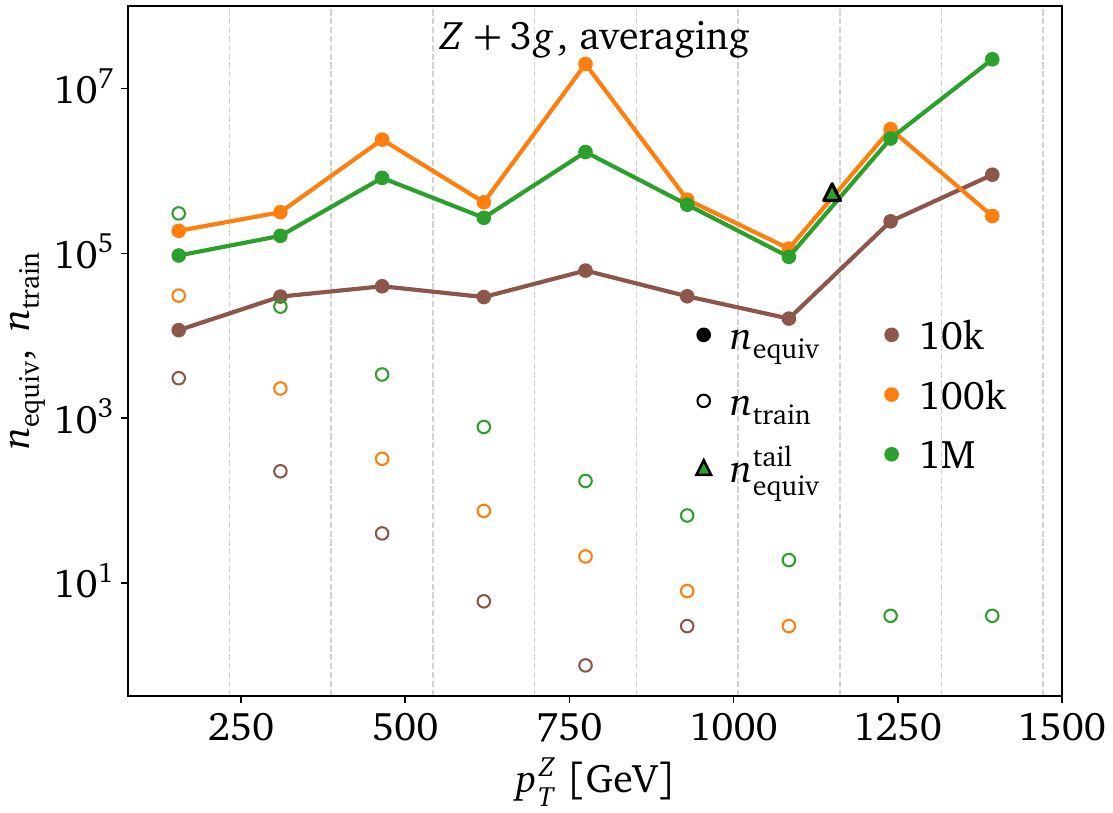}
\caption{Averaging amplification as a function of $p_T^Z$. For all bins, $\Nequiv$ is computed from the regular test dataset, for $p_T^Z \in [1100,1200]~\gev$ we also show the value computed from the targeted dataset with increased statistics.}
  \label{fig:zggg-nequiv-het}
\end{figure}

%----------------------------------------------------------
Finally, Figure~\ref{fig:zggg-nequiv-het} shows that the equivalent sample size remains larger than the available training statistics over most of the $p_T^Z$ spectrum, confirming the statistical amplification for the $Z+3g$ process.
%----------------------------------------------------------
%----------------------------------------------------------

\clearpage
%%%%%%%%%%%%%%%%%%%%%%%%%%%%%%%%%%%%%%%%%%%%%%%%%%%
\section{Log-Likelihood-Ratio Distributions}
\label{app:log-likelihood-ratios}

\begin{figure}[H]
  \includegraphics[width=0.49\textwidth]{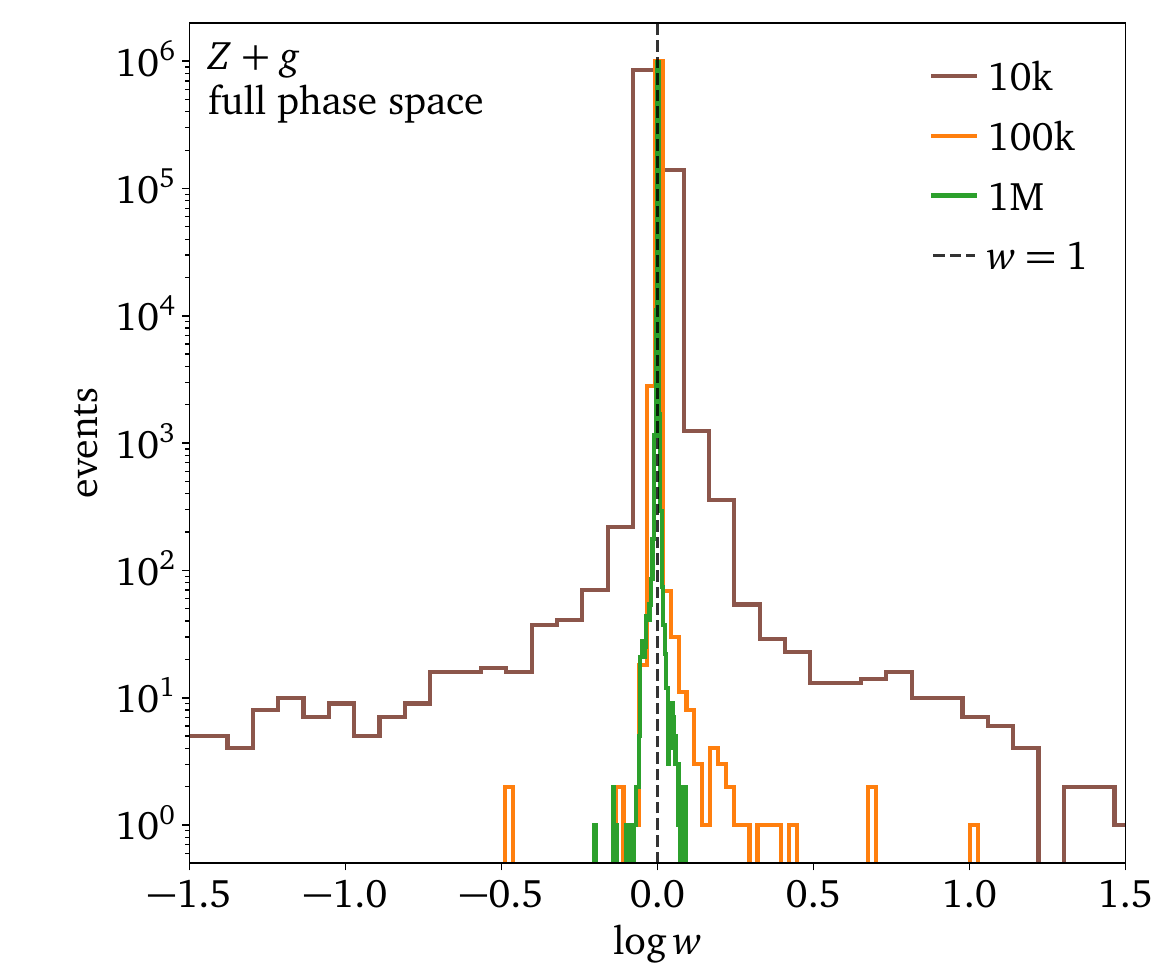}
  \includegraphics[width=0.49\textwidth]{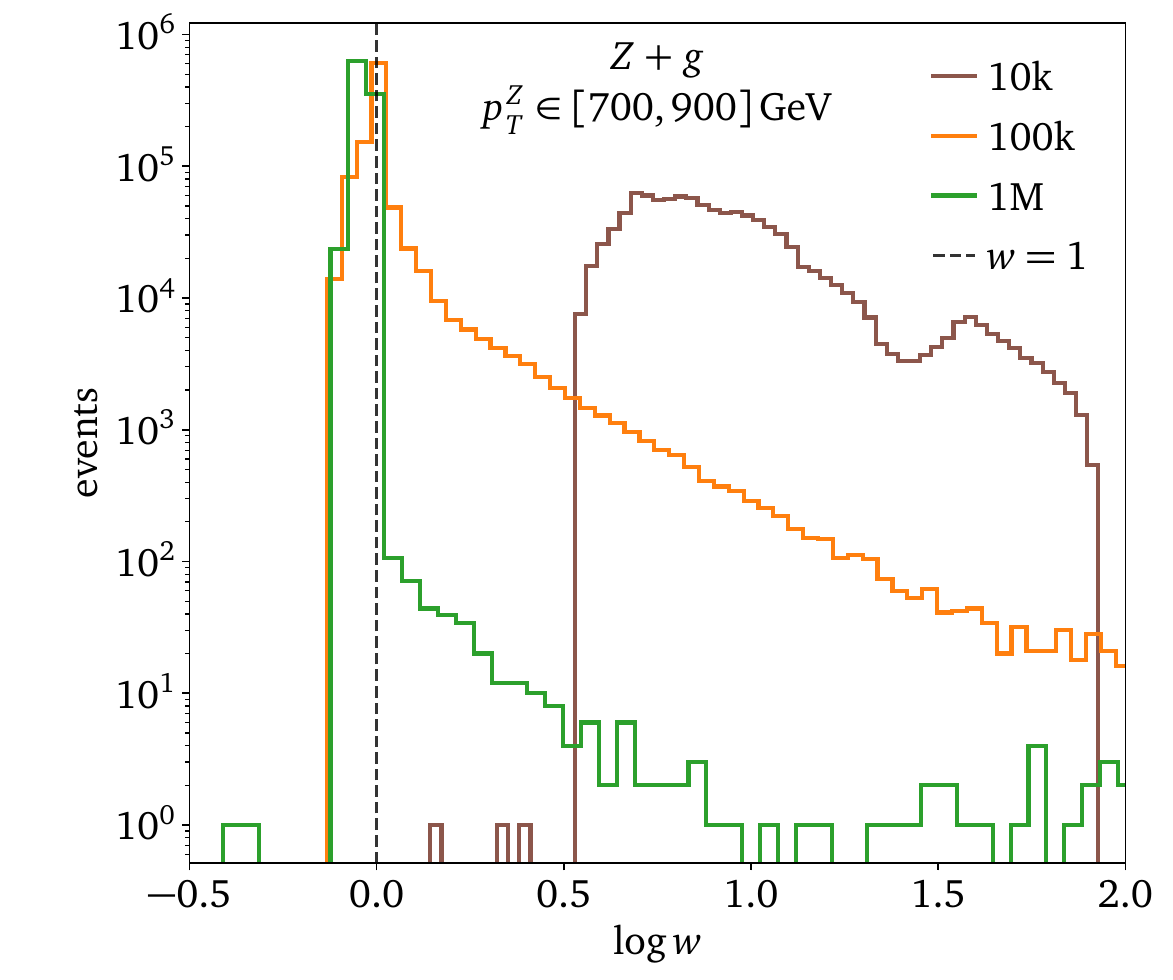}
\caption{
Log-likelihood ratio distributions for the $Z+g$ process for the three surrogate models. Left: original test sample. Right: targeted sample. The vertical dashed line marks $\log w=0$, corresponding to unit event weight ($w=1$).}
  \label{fig:zg-llr}
\end{figure}

\begin{figure}[H]
  \includegraphics[width=0.49\textwidth]{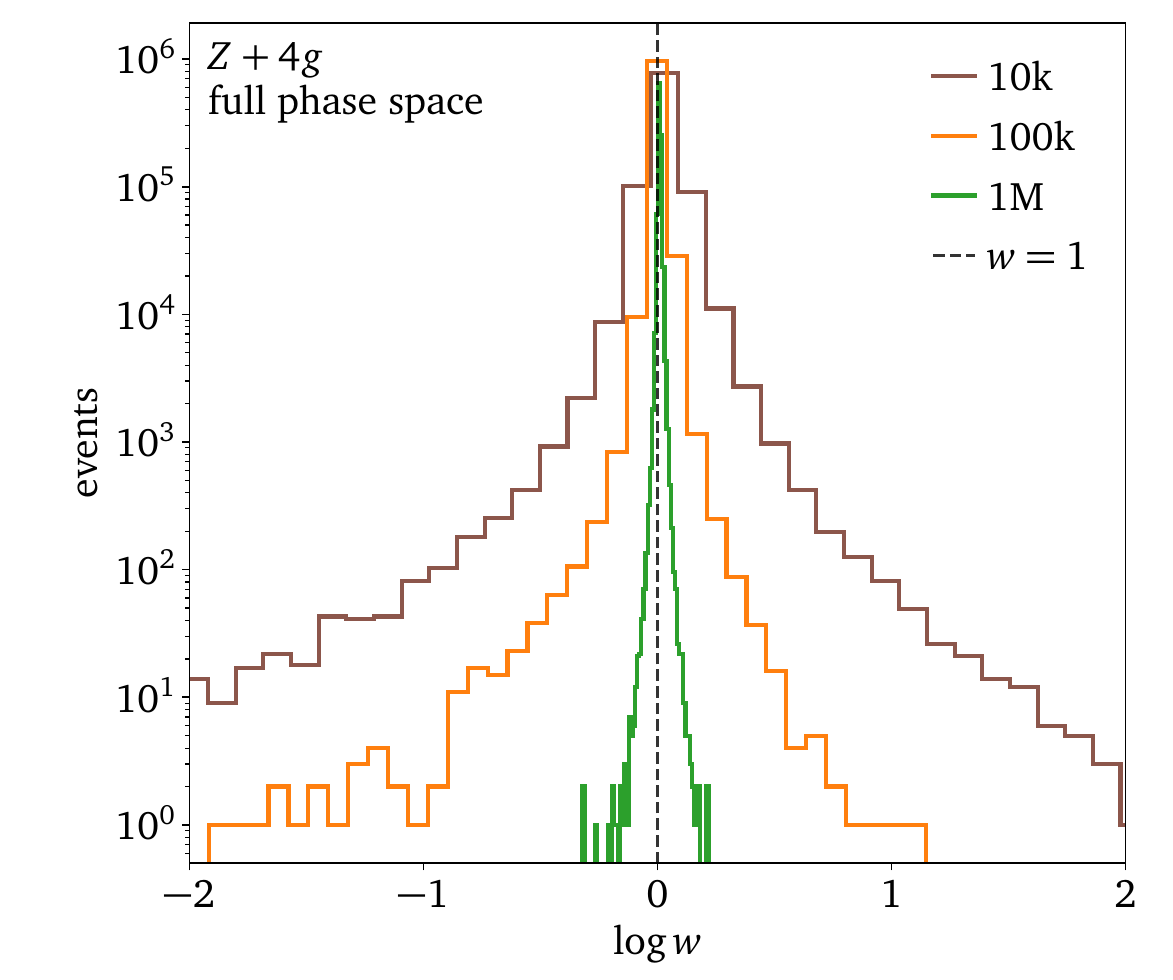}
  \includegraphics[width=0.49\textwidth]{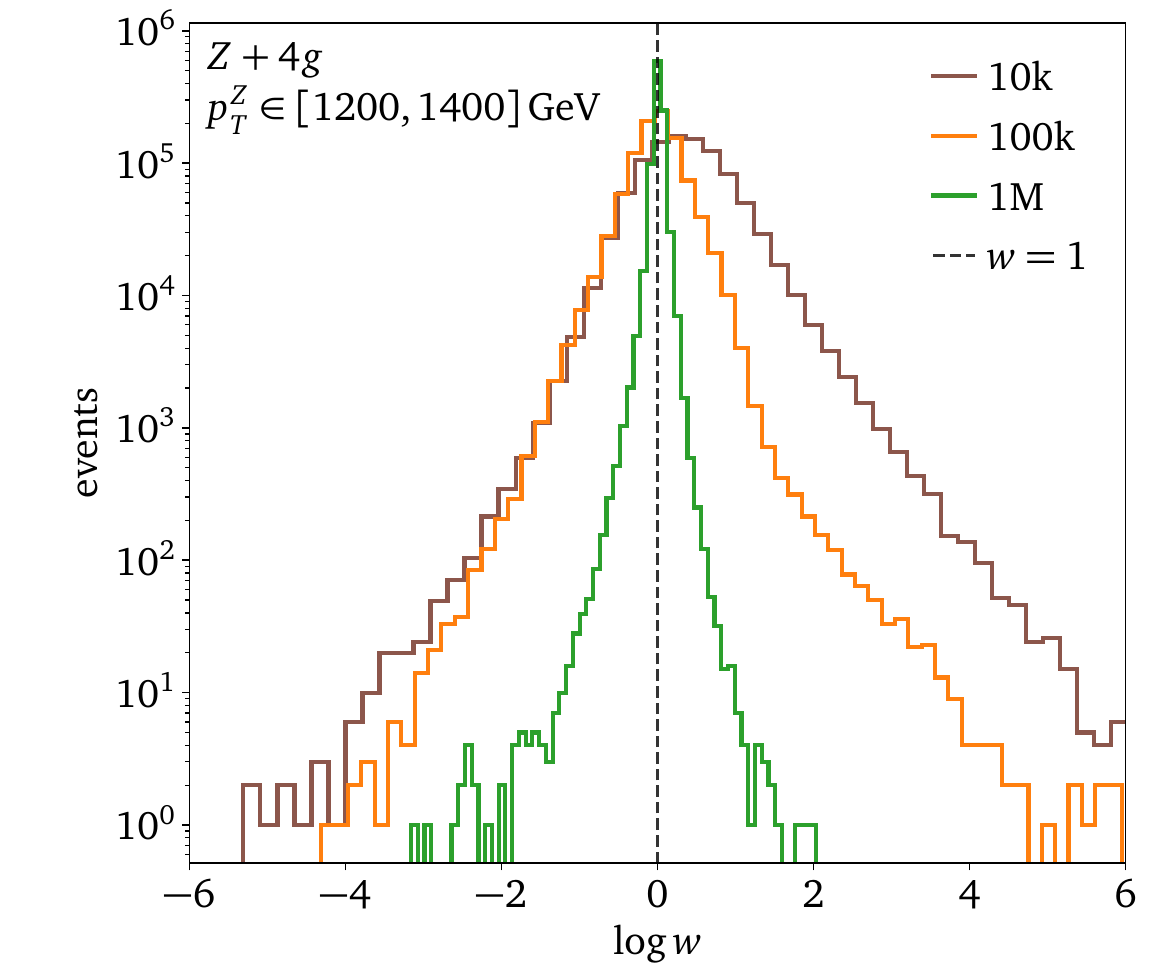}
\caption{
Log-likelihood ratio distributions for the $Z+4g$ process for the three surrogate models. Left: original test sample. Right: targeted sample. The vertical dashed line marks $\log w=0$, corresponding to unit event weight ($w=1$).}
  \label{fig:zgggg-llr}
\end{figure}

For completeness, we collect in Figures~\ref{fig:zg-llr}\,--\,\ref{fig:zgggg-llr} the log-likelihood ratio distributions for all considered processes, evaluated on the original test sample and the targeted sample for the three surrogate models. These distributions form the basis of the Kolmogorov--Smirnov statistic used to quantify the differential amplification. The vertical dashed line marks $w=1$ ($\log w=0$), corresponding to events for which the surrogate and the reference assign the same weight. As the training statistics increase, the distributions become progressively more concentrated around this value, reflecting the improved accuracy of the surrogate predictions.
%----------------------------------------------------------
%----------------------------------------------------------

%----------------------------------------------------------
\begin{figure}[H]
  \includegraphics[width=0.49\textwidth]{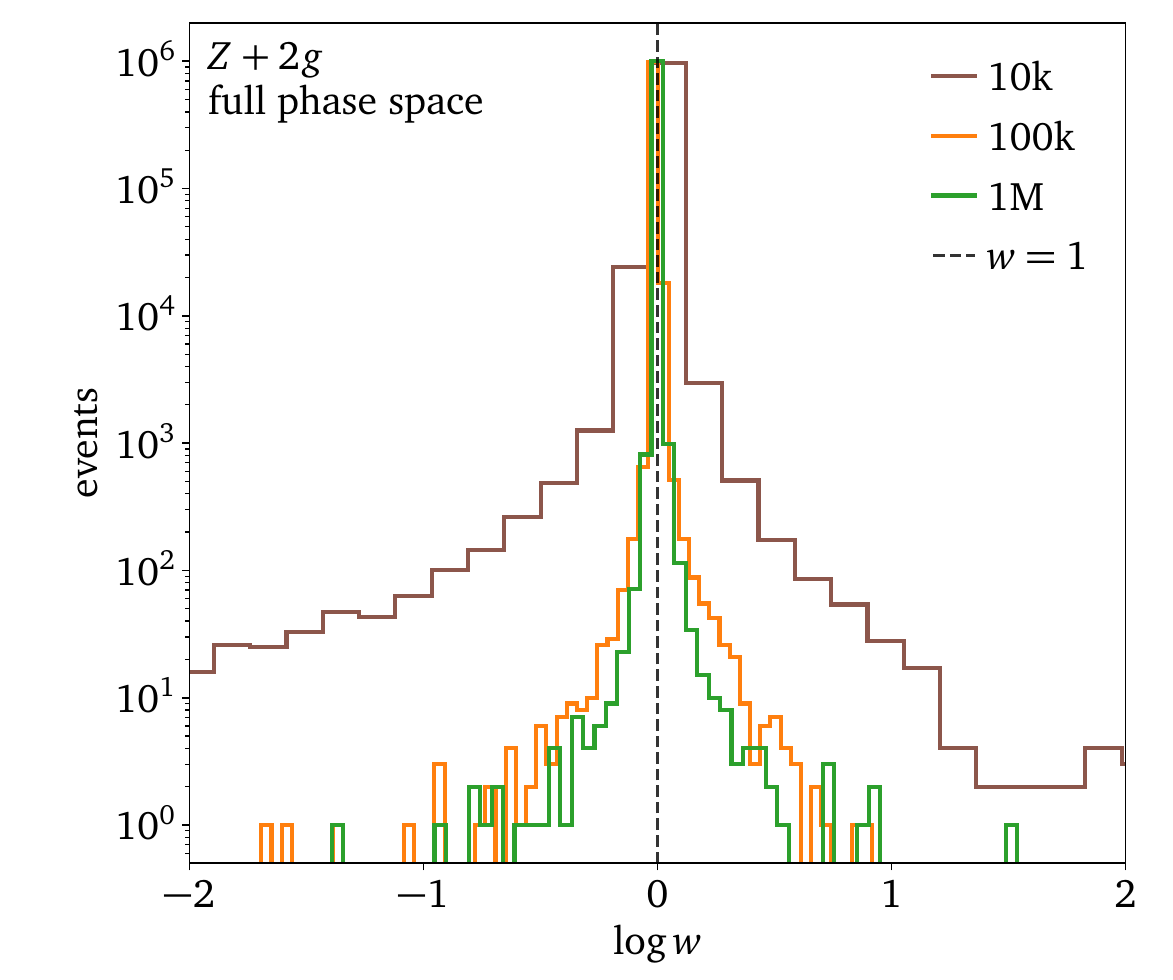}
  \includegraphics[width=0.49\textwidth]{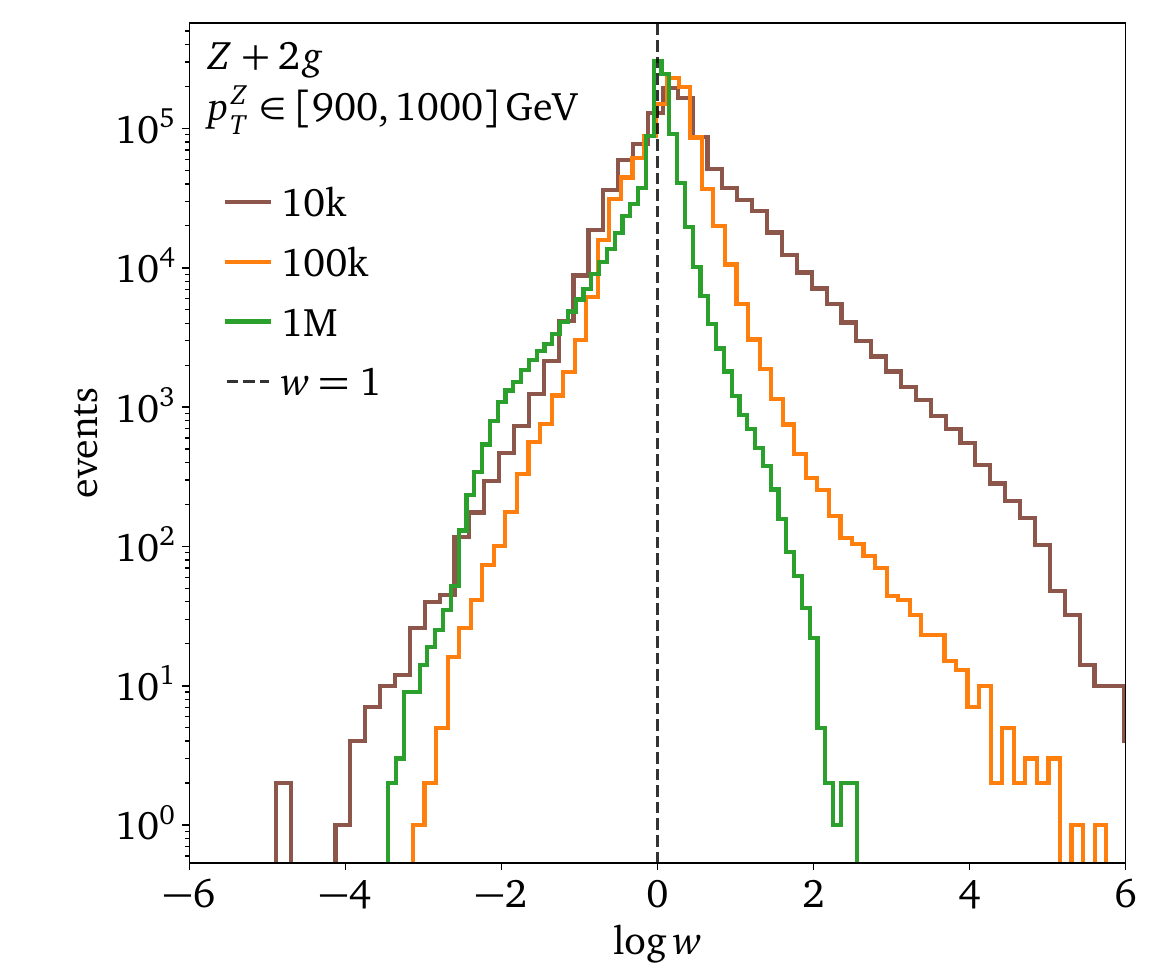}
\caption{
Log-likelihood ratio distributions for the $Z+2g$ process for the three surrogate models. Left: original test sample. Right: targeted sample. The vertical dashed line marks $\log w=0$, corresponding to unit event weight ($w=1$).}
  \label{fig:zgg-llr}
\end{figure}
%----------------------------------------------------------

%----------------------------------------------------------
\begin{figure}[H]
  \includegraphics[width=0.49\textwidth]{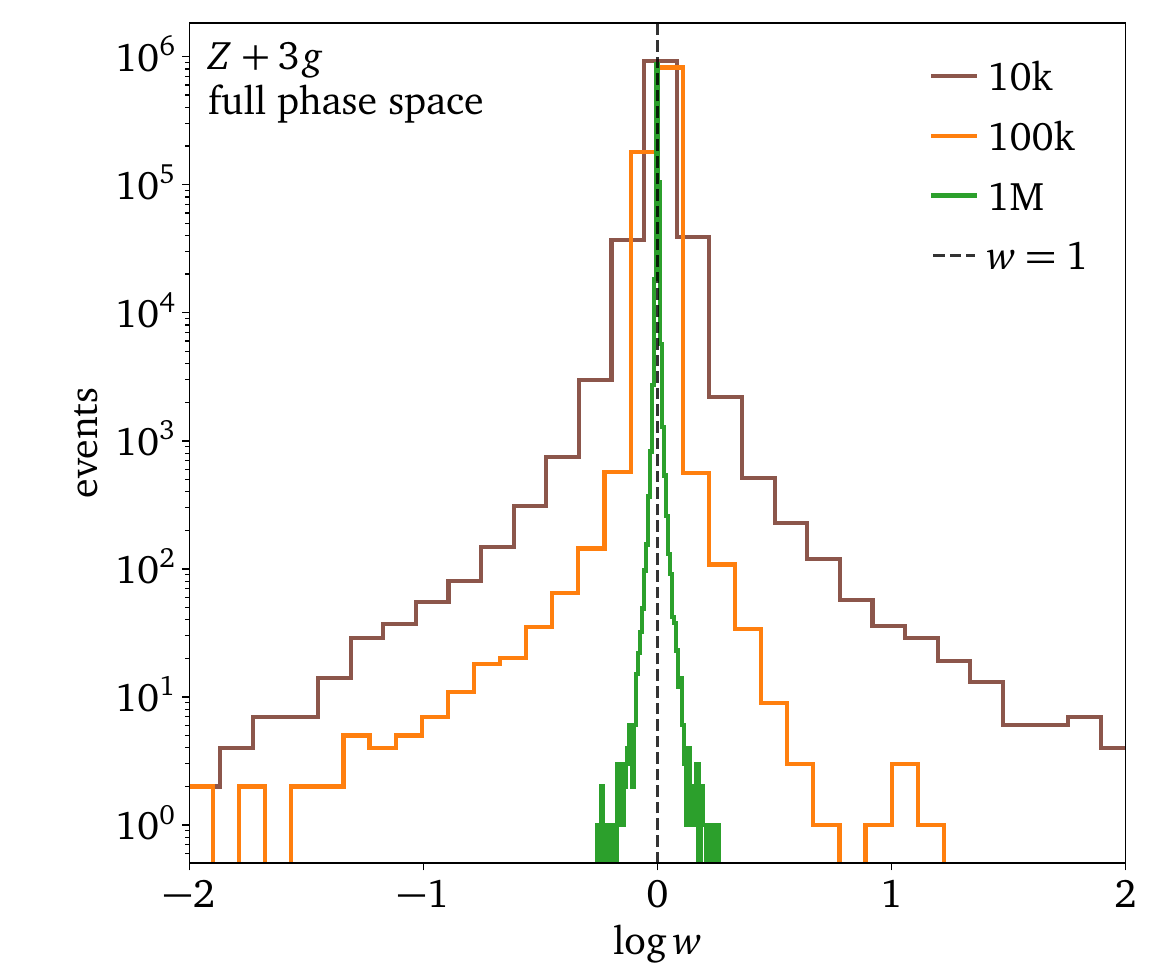}
  \includegraphics[width=0.49\textwidth]{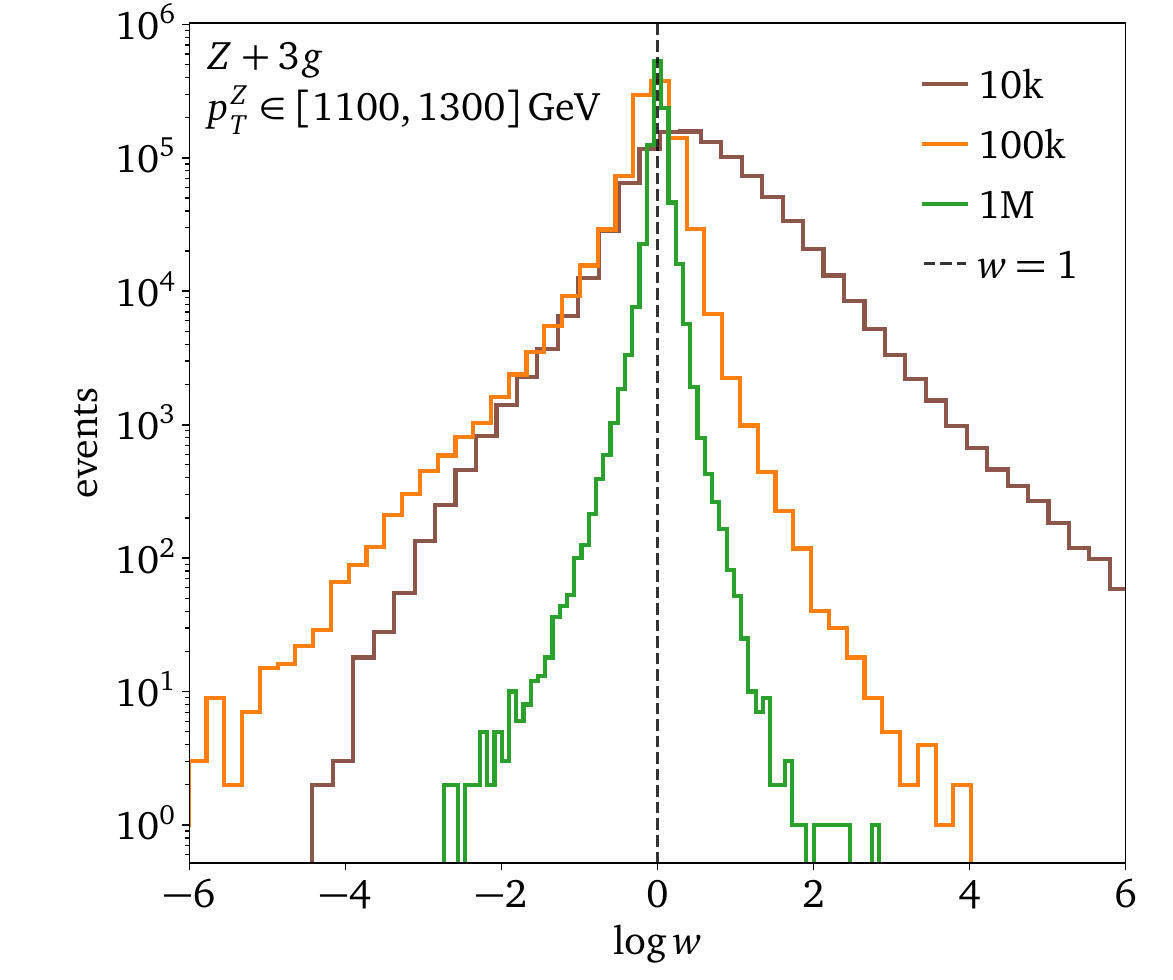}
\caption{
Log-likelihood ratio distributions for the $Z+3g$ process for the three surrogate models. Left: original test sample. Right: targeted sample. The vertical dashed line marks $\log w=0$, corresponding to unit event weight ($w=1$).}
  \label{fig:zggg-llr}
\end{figure}
%----------------------------------------------------------

%----------------------------------------------------------
%----------------------------------------------------------

\clearpage
%%%%%%%%%%%%%%%%%%%%%%%%%%%%%%%%%%%%%%%%%%%%%%%%%%
\bibliography{tilman,refs}

@article{Bieringer:2024nbc,
    author = "Bieringer, Sebastian and Diefenbacher, Sascha and Kasieczka, Gregor and Trabs, Mathias",
    title = "{Calibrating Bayesian generative machine learning for Bayesiamplification}",
    eprint = "2408.00838",
    archivePrefix = "arXiv",
    primaryClass = "cs.LG",
    doi = "10.1088/2632-2153/ad9136",
    journal = "Mach. Learn. Sci. Tech.",
    volume = "5",
    number = "4",
    pages = "045044",
    year = "2024"
}

@article{Heimel:2026hgp,
    author = "Heimel, Theo and Mattelaer, Olivier and Winterhalder, Ramon",
    title = "{MadSpace -- Event Generation for the Era of GPUs and ML}",
    eprint = "2602.06895",
    archivePrefix = "arXiv",
    primaryClass = "hep-ph",
    reportNumber = "MCNET-26-01, IRMP-CP3-26-04, TIF-UNIMI-2026-1",
    month = "2",
    year = "2026"
}

@article{Janssen:2025zke,
    author = "Jan{\ss}en, Timo and Poncelet, Rene and Schumann, Steffen",
    title = "{Sampling NNLO QCD phase space with normalizing flows}",
    eprint = "2505.13608",
    archivePrefix = "arXiv",
    primaryClass = "hep-ph",
    reportNumber = "IFJPAN-IV-2025-11, MCNET-25-11, COMETA-2025-22",
    doi = "10.1007/JHEP09(2025)194",
    journal = "JHEP",
    volume = "09",
    pages = "194",
    year = "2025"
}

@article{Bothmann:2025lwg,
    author = "Bothmann, E. and Jan{\ss}en, T. and Knobbe, M. and Schmitzer, B. and Sinz, F.",
    title = "{Efficient many-jet event generation with Flow Matching}",
    eprint = "2506.18987",
    archivePrefix = "arXiv",
    primaryClass = "hep-ph",
    reportNumber = "FERMILAB-PUB-25-0304-T, MCNET-25-15",
    month = "6",
    year = "2025"
}

@article{Deutschmann:2024lml,
    author = {Deutschmann, Nicolas and G\"otz, Niklas},
    title = "{Accelerating HEP simulations with Neural Importance Sampling}",
    eprint = "2401.09069",
    archivePrefix = "arXiv",
    primaryClass = "hep-ph",
    doi = "10.1007/JHEP03(2024)083",
    journal = "JHEP",
    volume = "03",
    pages = "083",
    year = "2024"
}

@article{Janssen:2023ahv,
    author = "Jan\ss{}en, Timo and Ma\^\i{}tre, Daniel and Schumann, Steffen and Siegert, Frank and Truong, Henry",
    title = "{Unweighting multijet event generation using factorisation-aware neural networks}",
    eprint = "2301.13562",
    archivePrefix = "arXiv",
    primaryClass = "hep-ph",
    reportNumber = "MCNET-23-01, IPPP/23/04",
    doi = "10.21468/SciPostPhys.15.3.107",
    journal = "SciPost Phys.",
    volume = "15",
    pages = "107",
    year = "2023"
}

@article{Maitre:2023dqz,
    author = "Ma\^\i{}tre, D. and Truong, H.",
    title = "{One-loop matrix element emulation with factorisation awareness}",
    eprint = "2302.04005",
    archivePrefix = "arXiv",
    primaryClass = "hep-ph",
    reportNumber = "IPPP/23/06",
	journal      = {JHEP},
	volume = {5},
	pages = {159},
    doi = "10.1007/JHEP05(2023)159",
    year = "2023"
}

@article{Danziger:2021eeg,
    author = "Danziger, Katharina and Jan\ss{}en, Timo and Schumann, Steffen and Siegert, Frank",
    title = "{Accelerating Monte Carlo event generation -- rejection sampling using neural network event-weight estimates}",
    eprint = "2109.11964",
    archivePrefix = "arXiv",
    primaryClass = "hep-ph",
    reportNumber = "MCNET-21-13",
    doi = "10.21468/SciPostPhys.12.5.164",
    journal = "SciPost Phys.",
    volume = "12",
    pages = "164",
    year = "2022"
}

@article{Verheyen:2022tov,
	author       = {Verheyen, Rob},
	title        = {{Event Generation and Density Estimation with Surjective Normalizing Flows}},
	year         = 2022,
	journal      = {SciPost Phys.},
	volume       = 13,
	number       = 3,
	pages        = {047},
	doi          = {10.21468/SciPostPhys.13.3.047},
	eprint       = {2205.01697},
	archiveprefix = {arXiv},
	primaryclass = {hep-ph}
}

@article{Aylett-Bullock:2021hmo,
	author       = {Aylett-Bullock, Joseph and Badger, Simon and Moodie, Ryan},
	title        = {{Optimising simulations for diphoton production at hadron colliders using amplitude neural networks}},
	year         = 2021,
	month        = 6,
	journal      = {JHEP},
	volume       = {08},
	pages        = {066},
	doi          = {10.1007/JHEP08(2021)066},
	eprint       = {2106.09474},
	archiveprefix = {arXiv},
	primaryclass = {hep-ph}
}

@article{Chen:2020nfb,
    author = "Chen, I-Kai and Klimek, Matthew D. and Perelstein, Maxim",
    title = "{Improved neural network Monte Carlo simulation}",
    eprint = "2009.07819",
    archivePrefix = "arXiv",
    primaryClass = "hep-ph",
    doi = "10.21468/SciPostPhys.10.1.023",
    journal = "SciPost Phys.",
    volume = "10",
    number = "1",
    pages = "023",
    year = "2021"
}

@article{Choi:2021sku,
	author       = {Choi, Suyong and Lim, Jae Hoon},
	title        = {{A Data-driven Event Generator for Hadron Colliders using Wasserstein Generative Adversarial Network}},
	year         = 2021,
	month        = 2,
	journal      = {J. Korean Phys. Soc.},
	volume       = 78,
	number       = 6,
	pages        = {482--489},
	doi          = {10.1007/s40042-021-00095-1},
	eprint       = {2102.11524},
	archiveprefix = {arXiv},
	primaryclass = {hep-ex}
}

@article{Maitre:2021uaa,
	author       = {Ma\^\i{}tre, Daniel and Truong, Henry},
	title        = {{A factorisation-aware Matrix element emulator}},
	year         = 2021,
	journal      = {JHEP},
	volume       = 11,
	pages        = {066},
	doi          = {10.1007/JHEP11(2021)066},
	eprint       = {2107.06625},
	archiveprefix = {arXiv},
	primaryclass = {hep-ph},
	reportnumber = {IPPP/21/11}
}

@article{Otten:2019hhl,
	author       = {Otten, Sydney and Caron, Sascha and de Swart, Wieske and van Beekveld, Melissa and Hendriks, Luc and van Leeuwen, Caspar and Podareanu, Damian and Ruiz de Austri, Roberto and Verheyen, Rob},
	title        = {{Event Generation and Statistical Sampling for Physics with Deep Generative Models and a Density Information Buffer}},
	year         = 2021,
	journal      = {Nature Commun.},
	volume       = 12,
	number       = 1,
	pages        = 2985,
	doi          = {10.1038/s41467-021-22616-z},
	eprint       = {1901.00875},
	archiveprefix = {arXiv},
	primaryclass = {hep-ph}
}

@article{Stienen:2020gns,
    author = "Stienen, Bob and Verheyen, Rob",
    title = "{Phase space sampling and inference from weighted events with autoregressive flows}",
    eprint = "2011.13445",
    archivePrefix = "arXiv",
    primaryClass = "hep-ph",
    doi = "10.21468/SciPostPhys.10.2.038",
    journal = "SciPost Phys.",
    volume = "10",
    number = "2",
    pages = "038",
    year = "2021"
}

@article{Alanazi:2020klf,
	author       = {Alanazi, Yasir and Sato, N. and Liu, Tianbo and Melnitchouk, W. and Kuchera, Michelle P. and Pritchard, Evan and Robertson, Michael and Strauss, Ryan and Velasco, Luisa and Li, Yaohang},
	title        = {{Simulation of electron-proton scattering events by a Feature-Augmented and Transformed Generative Adversarial Network (FAT-GAN)}},
	year         = 2020,
	month        = 1,
	eprint       = {2001.11103},
	archiveprefix = {arXiv},
	primaryclass = {hep-ph},
	reportnumber = {JLAB-THY-20-3136}
}

@article{Badger:2020uow,
	author       = {Badger, Simon and Bullock, Joseph},
	title        = {{Using neural networks for efficient evaluation of high multiplicity scattering amplitudes}},
	year         = 2020,
	journal      = {JHEP},
	volume       = {06},
	pages        = 114,
	doi          = {10.1007/JHEP06(2020)114},
	eprint       = {2002.07516},
	archiveprefix = {arXiv},
	primaryclass = {hep-ph},
	reportnumber = {IPPP/20/5}
}

@article{Bothmann:2020ywa,
    author = "Bothmann, Enrico and Jan\ss{}en, Timo and Knobbe, Max and Schmale, Tobias and Schumann, Steffen",
    title = "{Exploring phase space with Neural Importance Sampling}",
    eprint = "2001.05478",
    archivePrefix = "arXiv",
    primaryClass = "hep-ph",
    reportNumber = "MCNET-20-02, MCNET-20-01",
    doi = "10.21468/SciPostPhys.8.4.069",
    journal = "SciPost Phys.",
    volume = "8",
    number = "4",
    pages = "069",
    year = "2020"
}

@article{Gao:2020zvv,
	author       = {Gao, Christina and H\"oche, Stefan and Isaacson, Joshua and Krause, Claudius and Schulz, Holger},
	title        = {{Event Generation with Normalizing Flows}},
	year         = 2020,
	journal      = {Phys. Rev. D},
	volume       = 101,
	number       = 7,
	pages        = {076002},
	doi          = {10.1103/PhysRevD.101.076002},
	eprint       = {2001.10028},
	archiveprefix = {arXiv},
	primaryclass = {hep-ph},
	reportnumber = {FERMILAB-PUB-20-009-SCD-T, MCNET-20-03}
}

@article{Gao:2020vdv,
    author = "Gao, Christina and Isaacson, Joshua and Krause, Claudius",
    title = "{i-flow: High-dimensional Integration and Sampling with Normalizing Flows}",
    eprint = "2001.05486",
    archivePrefix = "arXiv",
    primaryClass = "physics.comp-ph",
    reportNumber = "FERMILAB-PUB-20-010-T",
    doi = "10.1088/2632-2153/abab62",
    journal = "Mach. Learn. Sci. Tech.",
    volume = "1",
    number = "4",
    pages = "045023",
    year = "2020"
}

@article{Klimek:2018mza,
    author = "Klimek, Matthew D. and Perelstein, Maxim",
    title = "{Neural Network-Based Approach to Phase Space Integration}",
    eprint = "1810.11509",
    archivePrefix = "arXiv",
    primaryClass = "hep-ph",
    doi = "10.21468/SciPostPhys.9.4.053",
    journal = "SciPost Phys.",
    volume = "9",
    pages = "053",
    year = "2020"
}

@article{Bishara:2019iwh,
    author = "Bishara, Fady and Montull, Marc",
    title = "{Machine learning amplitudes for faster event generation}",
    eprint = "1912.11055",
    archivePrefix = "arXiv",
    primaryClass = "hep-ph",
    reportNumber = "DESY 19-232, DESY-19-232",
    doi = "10.1103/PhysRevD.107.L071901",
    journal = "Phys. Rev. D",
    volume = "107",
    number = "7",
    pages = "L071901",
    year = "2023"
}

@article{DiSipio:2019imz,
	author       = {Di Sipio, Riccardo and Faucci Giannelli, Michele and Ketabchi Haghighat, Sana and Palazzo, Serena},
	title        = {{DijetGAN: A Generative-Adversarial Network Approach for the Simulation of QCD Dijet Events at the LHC}},
	year         = 2019,
	journal      = {JHEP},
	volume       = {08},
	pages        = 110,
	doi          = {10.1007/JHEP08(2019)110},
	eprint       = {1903.02433},
	archiveprefix = {arXiv},
	primaryclass = {hep-ex},
	slaccitation = {%%CITATION = ARXIV:1903.02433;%%}
}

@article{Hashemi:2019fkn,
	author       = {Hashemi, Bobak and Amin, Nick and Datta, Kaustuv and Olivito, Dominick and Pierini, Maurizio},
	title        = {{LHC analysis-specific datasets with Generative Adversarial Networks}},
	year         = 2019,
	eprint       = {1901.05282},
	archiveprefix = {arXiv},
	primaryclass = {hep-ex},
	slaccitation = {%%CITATION = ARXIV:1901.05282;%%}
}

@article{Bendavid:2017zhk,
	author       = {Bendavid, Joshua},
	title        = {{Efficient Monte Carlo Integration Using Boosted Decision Trees and Generative Deep Neural Networks}},
	year         = 2017,
	month        = 6,
	eprint       = {1707.00028},
	archiveprefix = {arXiv},
	primaryclass = {hep-ph}
}

@article{Alwall:2014hca,
	author       = {Alwall, J. and Frederix, R. and Frixione, S. and Hirschi, V. and Maltoni, F. and Mattelaer, O. and Shao, H. -S. and Stelzer, T. and Torrielli, P. and Zaro, M.},
	title        = {{The automated computation of tree-level and next-to-leading order differential cross sections, and their matching to parton shower simulations}},
	year         = 2014,
	journal      = {JHEP},
	volume       = {07},
	pages        = {079},
	doi          = {10.1007/JHEP07(2014)079},
	eprint       = {1405.0301},
	archiveprefix = {arXiv},
	primaryclass = {hep-ph},
	reportnumber = {CERN-PH-TH-2014-064, CP3-14-18, LPN14-066, MCNET-14-09, ZU-TH-14-14}
}

@article{Breso-Pla:2024pda,
    author = "Bres{\'o}-Pla, V{\'\i}ctor and Heinrich, Gudrun and Magerya, Vitaly and Olsson, Anton",
    title = "{Interpolating amplitudes}",
    eprint = "2412.09534",
    archivePrefix = "arXiv",
    primaryClass = "hep-ph",
    reportNumber = "CERN-TH-2024-211, KA-TP-23-2024, P3H-24-092",
    doi = "10.21468/SciPostPhys.19.5.123",
    journal = "SciPost Phys.",
    volume = "19",
    number = "5",
    pages = "123",
    year = "2025"
}

@article{Herrmann:2025nnz,
    author = "Herrmann, Tim and Jan{\ss}en, Timo and Schenker, Mathis and Schumann, Steffen and Siegert, Frank",
    title = "{Accelerating multijet-merged event generation with neural network matrix element surrogates}",
    eprint = "2506.06203",
    archivePrefix = "arXiv",
    primaryClass = "hep-ph",
    reportNumber = "MCNET-25-12",
    doi = "10.21468/SciPostPhys.20.3.071",
    journal = "SciPost Phys.",
    volume = "20",
    pages = "071",
    year = "2026"
}

@article{Beccatini:2025tpk,
    author = "Beccatini, Luca and Maltoni, Fabio and Mattelaer, Olivier and Winterhalder, Ramon",
    title = "{Amplitude Surrogates for Multi-Jet Processes}",
    eprint = "2512.11036",
    archivePrefix = "arXiv",
    primaryClass = "hep-ph",
    reportNumber = "TIF-UNIMI-2025-26, IRMP-CP3-25-44",
    month = "12",
    year = "2025"
}

@article{Bahl:2026jvt,
    author = "Bahl, Henning and Bres{\'o}-Pla, Victor and Butter, Anja and Ramirez, Joaqu{\'\i}n Iturriza",
    title = "{Scaling laws for amplitude surrogates}",
    eprint = "2601.13308",
    archivePrefix = "arXiv",
    primaryClass = "hep-ph",
    month = "1",
    year = "2026"
}

@article{Favaro:2026amj,
    author = "Favaro, Luigi and Plehn, Tilman and Qu, Huilin and Spinner, Jonas",
    title = "{Virtues and Vices of Equivariant Transformers}",
    eprint = "2608.02735",
    archivePrefix = "arXiv",
    primaryClass = "hep-ph",
    reportNumber = "IRMP-CP3-26-23",
    month = "8",
    year = "2026"
}

@article{Heimel:2026cxh,
    author = "Heimel, Theo and Plehn, Tilman and Revelli, Rebecca and Vent, Sophia and Winterhalder, Ramon",
    title = "{Neural Control Variates at LO and NLO}",
    eprint = "2607.23591",
    archivePrefix = "arXiv",
    primaryClass = "hep-ph",
    reportNumber = "TIF-UNIMI-2026-9",
    month = "7",
    year = "2026"
}

@article{Dubey:2026cts,
    author = {Dubey, Suprio and Bahl, Henning and Butter, Anja and Hesser, J{\"u}rgen and Plehn, Tilman},
    title = "{Local Conformal Predictions for Calibrated Surrogates}",
    eprint = "2607.01354",
    archivePrefix = "arXiv",
    primaryClass = "hep-ph",
    month = "7",
    year = "2026"
}

@article{DeCrescenzo:2026tsp,
    author = "De Crescenzo, Giovanni and Villadamigo, Javier Mari{\~n}o and Elmer, Nina and Heimel, Theo and Plehn, Tilman and Winterhalder, Ramon and Zaro, Marco",
    title = "{MadNIS at NLO}",
    eprint = "2603.22407",
    archivePrefix = "arXiv",
    primaryClass = "hep-ph",
    reportNumber = "TIF-UNIMI-2026-2, IRMP-CP3-26-06, MCNET-26-04",
    month = "3",
    year = "2026"
}

@article{Bahl:2026qaf,
    author = "Bahl, Henning and Braun, Jens and Heinrich, Gudrun and Plehn, Tilman and Revelli, Rebecca",
    title = "{How to Trust Learned Loop Amplitudes}",
    eprint = "2601.00950",
    archivePrefix = "arXiv",
    primaryClass = "hep-ph",
    month = "1",
    year = "2026"
}

@article{Petitjean:2025zjf,
    author = {Petitjean, Antoine and Plehn, Tilman and Spinner, Jonas and K{\"o}the, Ullrich},
    title = "{Economical Jet Taggers -- Equivariant, Slim, and Quantized}",
    eprint = "2512.17011",
    archivePrefix = "arXiv",
    primaryClass = "hep-ph",
    reportNumber = "IPPP/25/93",
    month = "12",
    year = "2025"
}

@article{Bahl:2025ryd,
    author = "Bahl, Henning and Diefenbacher, Sascha and Elmer, Nina and Plehn, Tilman and Spinner, Jonas",
    title = "{Forecasting generative amplification}",
    eprint = "2509.08048",
    archivePrefix = "arXiv",
    primaryClass = "hep-ph",
    doi = "10.21468/SciPostPhys.20.5.150",
    journal = "SciPost Phys.",
    volume = "20",
    number = "5",
    pages = "150",
    year = "2026"
}

@article{Villadamigo:2025our,
    author = "Villadamigo, Javier Mari{\~n}o and Frederix, Rikkert and Plehn, Tilman and Vitos, Timea and Winterhalder, Ramon",
    title = "{FASTColor -- Full-color Amplitude Surrogate Toolkit for QCD}",
    eprint = "2509.07068",
    archivePrefix = "arXiv",
    primaryClass = "hep-ph",
    reportNumber = "TIF-UNIMI-2025-18",
    doi = "10.21468/SciPostPhys.21.1.001",
    journal = "SciPost Phys.",
    volume = "21",
    pages = "001",
    year = "2026"
}

@article{Bahl:2025xvx,
    author = "Bahl, Henning and Elmer, Nina and Plehn, Tilman and Winterhalder, Ramon",
    title = "{Amplitude Uncertainties Everywhere All at Once}",
    eprint = "2509.00155",
    archivePrefix = "arXiv",
    primaryClass = "hep-ph",
    reportNumber = "TIF-UNIMI-2025-17",
    doi = "10.21468/SciPostPhys.20.3.083",
    journal = "SciPost Phys.",
    volume = "20",
    pages = "083",
    year = "2026"
}

@article{Favaro:2025pgz,
    author = "Favaro, Luigi and Gerhartz, Gerrit and Hamprecht, Fred A. and Lippmann, Peter and Pitz, Sebastian and Plehn, Tilman and Qu, Huilin and Spinner, Jonas",
    title = "{Lorentz-Equivariance without Limitations}",
    eprint = "2508.14898",
    archivePrefix = "arXiv",
    primaryClass = "hep-ph",
    month = "8",
    year = "2025"
}

@article{Bahl:2024gyt,
    author = "Bahl, Henning and Elmer, Nina and Favaro, Luigi and Haussmann, Manuel and Plehn, Tilman and Winterhalder, Ramon",
    title = "{Accurate surrogate amplitudes with calibrated uncertainties}",
    eprint = "2412.12069",
    archivePrefix = "arXiv",
    primaryClass = "hep-ph",
    doi = "10.21468/SciPostPhysCore.8.4.073",
    journal = "SciPost Phys. Core",
    volume = "8",
    pages = "073",
    year = "2025"
}

@article{Brehmer:2024yqw,
    author = "Brehmer, Johann and Bres{\'o}, V{\'\i}ctor and de Haan, Pim and Plehn, Tilman and Qu, Huilin and Spinner, Jonas and Thaler, Jesse",
    title = "{A Lorentz-equivariant transformer for all of the LHC}",
    eprint = "2411.00446",
    archivePrefix = "arXiv",
    primaryClass = "hep-ph",
    reportNumber = "MIT-CTP/5802",
    doi = "10.21468/SciPostPhys.19.4.108",
    journal = "SciPost Phys.",
    volume = "19",
    number = "4",
    pages = "108",
    year = "2025"
}

@article{Heimel:2024wph,
    author = "Heimel, Theo and Mattelaer, Olivier and Plehn, Tilman and Winterhalder, Ramon",
    title = "{Differentiable MadNIS-Lite}",
    eprint = "2408.01486",
    archivePrefix = "arXiv",
    primaryClass = "hep-ph",
    reportNumber = "IRMP-CP3-24-23",
    doi = "10.21468/SciPostPhys.18.1.017",
    journal = "SciPost Phys.",
    volume = "18",
    number = "1",
    pages = "017",
    year = "2025"
}

@article{Heimel:2023ngj,
    author = "Heimel, Theo and Huetsch, Nathan and Maltoni, Fabio and Mattelaer, Olivier and Plehn, Tilman and Winterhalder, Ramon",
    title = "{The MadNIS reloaded}",
    eprint = "2311.01548",
    archivePrefix = "arXiv",
    primaryClass = "hep-ph",
    reportNumber = "IRMP-CP3-23-56, MCNET-23-12",
    doi = "10.21468/SciPostPhys.17.1.023",
    journal = "SciPost Phys.",
    volume = "17",
    number = "1",
    pages = "023",
    year = "2024"
}

@article{Heimel:2023mvw,
    author = "Heimel, Theo and Huetsch, Nathan and Winterhalder, Ramon and Plehn, Tilman and Butter, Anja",
    title = "{Precision-machine learning for the matrix element method}",
    eprint = "2310.07752",
    archivePrefix = "arXiv",
    primaryClass = "hep-ph",
    reportNumber = "IRMP-CP3-23-55",
    doi = "10.21468/SciPostPhys.17.5.129",
    journal = "SciPost Phys.",
    volume = "17",
    number = "5",
    pages = "129",
    year = "2024"
}

@article{Butter:2023fov,
    author = "Butter, Anja and Huetsch, Nathan and Palacios Schweitzer, Sofia and Plehn, Tilman and Sorrenson, Peter and Spinner, Jonas",
    title = "{Jet diffusion versus JetGPT {\textendash} Modern networks for the LHC}",
    eprint = "2305.10475",
    archivePrefix = "arXiv",
    primaryClass = "hep-ph",
    doi = "10.21468/SciPostPhysCore.8.1.026",
    journal = "SciPost Phys. Core",
    volume = "8",
    pages = "026",
    year = "2025"
}

@article{Heimel:2022wyj,
    author = "Heimel, Theo and Winterhalder, Ramon and Butter, Anja and Isaacson, Joshua and Krause, Claudius and Maltoni, Fabio and Mattelaer, Olivier and Plehn, Tilman",
    title = "{MadNIS - Neural multi-channel importance sampling}",
    eprint = "2212.06172",
    archivePrefix = "arXiv",
    primaryClass = "hep-ph",
    reportNumber = "IRMP-CP3-22-56, MCNET-22-22, FERMILAB-PUB-22-915-T",
    doi = "10.21468/SciPostPhys.15.4.141",
    journal = "SciPost Phys.",
    volume = "15",
    number = "4",
    pages = "141",
    year = "2023"
}

@article{Plehn:2022ftl,
    author = "Plehn, Tilman and Butter, Anja and Dillon, Barry and Heimel, Theo and Krause, Claudius and Winterhalder, Ramon",
    title = "{Modern Machine Learning for LHC Physicists}",
    eprint = "2211.01421",
    archivePrefix = "arXiv",
    primaryClass = "hep-ph",
    month = "11",
    year = "2022"
}

@article{Badger:2022hwf,
    author = "Badger, Simon and Butter, Anja and Luchmann, Michel and Pitz, Sebastian and Plehn, Tilman",
    title = "{Loop amplitudes from precision networks}",
    eprint = "2206.14831",
    archivePrefix = "arXiv",
    primaryClass = "hep-ph",
    doi = "10.21468/SciPostPhysCore.6.2.034",
    journal = "SciPost Phys. Core",
    volume = "6",
    pages = "034",
    year = "2023"
}

@article{Butter:2022rso,
    author = "Badger, Simon and others",
    editor = "Butter, Anja and Plehn, Tilman and Schumann, Steffen",
    title = "{Machine learning and LHC event generation}",
    eprint = "2203.07460",
    archivePrefix = "arXiv",
    primaryClass = "hep-ph",
    reportNumber = "FERMILAB-PUB-22-126-T",
    doi = "10.21468/SciPostPhys.14.4.079",
    journal = "SciPost Phys.",
    volume = "14",
    number = "4",
    pages = "079",
    year = "2023"
}

@article{Bieringer:2022cbs,
    author = "Bieringer, Sebastian and Butter, Anja and Diefenbacher, Sascha and Eren, Engin and Gaede, Frank and Hundhausen, Daniel and Kasieczka, Gregor and Nachman, Benjamin and Plehn, Tilman and Trabs, Mathias",
    title = "{Calomplification {\textemdash} the power of generative calorimeter models}",
    eprint = "2202.07352",
    archivePrefix = "arXiv",
    primaryClass = "hep-ph",
    reportNumber = "DESY-22-031",
    doi = "10.1088/1748-0221/17/09/P09028",
    journal = "JINST",
    volume = "17",
    number = "09",
    pages = "P09028",
    year = "2022"
}

@article{Winterhalder:2021ngy,
    author = "Winterhalder, Ramon and Magerya, Vitaly and Villa, Emilio and Jones, Stephen P. and Kerner, Matthias and Butter, Anja and Heinrich, Gudrun and Plehn, Tilman",
    title = "{Targeting multi-loop integrals with neural networks}",
    eprint = "2112.09145",
    archivePrefix = "arXiv",
    primaryClass = "hep-ph",
    reportNumber = "CP3-21-65, KA-TP-29-2021, P3H-21-105",
    doi = "10.21468/SciPostPhys.12.4.129",
    journal = "SciPost Phys.",
    volume = "12",
    number = "4",
    pages = "129",
    year = "2022"
}

@article{Butter:2021csz,
    author = "Butter, Anja and Heimel, Theo and Hummerich, Sander and Krebs, Tobias and Plehn, Tilman and Rousselot, Armand and Vent, Sophia",
    title = "{Generative networks for precision enthusiasts}",
    eprint = "2110.13632",
    archivePrefix = "arXiv",
    primaryClass = "hep-ph",
    doi = "10.21468/SciPostPhys.14.4.078",
    journal = "SciPost Phys.",
    volume = "14",
    number = "4",
    pages = "078",
    year = "2023"
}

@article{Bellagente:2021yyh,
    author = "Bellagente, Marco and Haussmann, Manuel and Luchmann, Michel and Plehn, Tilman",
    title = "{Understanding Event-Generation Networks via Uncertainties}",
    eprint = "2104.04543",
    archivePrefix = "arXiv",
    primaryClass = "hep-ph",
    doi = "10.21468/SciPostPhys.13.1.003",
    journal = "SciPost Phys.",
    volume = "13",
    number = "1",
    pages = "003",
    year = "2022"
}

@article{Butter:2020qhk,
    author = "Butter, Anja and Diefenbacher, Sascha and Kasieczka, Gregor and Nachman, Benjamin and Plehn, Tilman",
    title = "{GANplifying event samples}",
    eprint = "2008.06545",
    archivePrefix = "arXiv",
    primaryClass = "hep-ph",
    doi = "10.21468/SciPostPhys.10.6.139",
    journal = "SciPost Phys.",
    volume = "10",
    number = "6",
    pages = "139",
    year = "2021"
}

@article{Butter:2019cae,
    author = "Butter, Anja and Plehn, Tilman and Winterhalder, Ramon",
    title = "{How to GAN LHC Events}",
    eprint = "1907.03764",
    archivePrefix = "arXiv",
    primaryClass = "hep-ph",
    doi = "10.21468/SciPostPhys.7.6.075",
    journal = "SciPost Phys.",
    volume = "7",
    number = "6",
    pages = "075",
    year = "2019"
}
\end{document}